\documentclass[journal]{IEEEtran}

\usepackage{amsmath,graphicx}
\usepackage{algorithm,algorithmic}
\usepackage{cite,amssymb,amsfonts,color,textcomp}
\usepackage{bm}
\usepackage{balance}
\usepackage{booktabs}
\usepackage{amsthm}
\usepackage{setspace}

\makeatletter
\newcommand*{\rom}[1]{\expandafter\@slowromancap\romannumeral #1@}
\makeatother

\def\abf{{\bf a}}

\def\cbf{{\bf c}}

\def\fbf{{\bf f}}
\def\gbf{{\bf g}}
\def\hbf{{\bf h}}

\def\qbf{{\bf q}}

\def\wbf{{\bf w}}
\def\xbf{{\bf x}}
\def\ybf{{\bf y}}
\def\zbf{{\bf z}}

\def\xbf{{\bf x}}
\def\ybf{{\bf y}}

\def\Hbf{{\bf H}}
\def\Ibf{{\bf I}}

\def\Rbf{{\bf R}}

\def\Gc{{\cal G}}

\def\Ic{{\cal I}}

\def\Kc{{\cal K}}

\def\Pc{{\cal P}}
\def\Qc{{\cal Q}}

\def\Sc{{\cal S}}
\def\Tc{{\cal T}}
\def\Uc{{\cal U}}

\def\Yc{{\cal Y}}
\def\Zc{{\cal Z}}

\def\ie{{\it i.e.,\ \/}}
\def\nn{\nonumber}

\def\hhbf{{\hat{\bf h}}}

\theoremstyle{definition}

\newtheorem{remark}{Remark}

\newtheorem{definition}{Definition}

\newenvironment{mylist}%
{\begin{list}{}%
    {%
      \setlength{\itemindent}{-5pt}%
      \setlength{\leftmargin}{12pt}%
      \setlength{\parsep}{\parskip}
      \setlength{\labelsep}{5pt}
      \setlength{\itemsep}{2pt}}}%
  {\end{list}}

\begin{document}

\title{Fast Group Scheduling for Downlink Large-Scale Multi-Group Multicast Beamforming}

\author{Chong~Zhang,~\IEEEmembership{Student~Member,~IEEE,}
        Min~Dong,~\IEEEmembership{Fellow,~IEEE,}
        Ben~Liang,~\IEEEmembership{Fellow,~IEEE,}
        Ali~Afana,~\IEEEmembership{Member,~IEEE,}
        and~Yahia~Ahmed%
        \thanks{Chong Zhang and Ben Liang are with the Department of Electrical and Computer Engineering, University of Toronto, Canada (e-mail: \{chongzhang,liang\}@ece.utoronto.ca). Min Dong is with the Department of Electrical, Computer, and Software Engineering, Ontario Tech University, Canada (e-mail: min.dong@ontariotechu.ca). Ali Afana and Yahia Ahmed are with Ericsson Canada (e-mail: \{ali.afana, yahia.ahmed\}@ericsson.com).}
}%

\maketitle

\begin{abstract}
Next-generation wireless networks need to  handle massive user access effectively. This paper addresses the problem of  joint group scheduling and multicast beamforming for downlink transmission with many active user groups. Aiming to maximize the minimum user throughput, we propose a three-phase approach to tackle this difficult joint optimization problem efficiently.
In Phase 1, we utilize the  optimal multicast beamforming structure obtained recently  to find the group-channel directions  for all groups.
We propose two low-complexity group scheduling algorithms in Phase 2, which
determine the subset of groups in each time slot sequentially and the total number of time slots required for all groups.  The first algorithm measures the level of spatial separation among groups and selects the dissimilar groups that maximize the minimum user rate into the same time slot. In contrast, the second algorithm first identifies the spatially correlated  groups via a learning-based clustering method based on the group-channel directions, and then separates spatially similar groups into different time slots.
Finally, the multicast beamformers for the scheduled groups are obtained in each time slot by a computationally efficient method.
Simulation results show that our proposed scheduling methods can effectively capture the level of spatial separation among groups to  improve the minimum user throughput over the conventional approach that serves all groups in a single time slot or one group per time slot, and can be executed with low computational complexity. \end{abstract}


\section{Introduction}
\label{sec:intro}

Content distribution through wireless multicasting  has become
increasingly popular in the fast growing wireless services
and applications \cite{Araniti&etal:Netw2017}. With unprecedented massive user access for content sharing and distribution, future wireless networks need to provide intelligent transmission and effective resource management to deliver the massive wireless traffic with high  efficiency.
 For downlink data distribution, multicast beamforming is an efficient transmission technique to deliver common messages to multiple groups of users  simultaneously with improved power and spectrum efficiency.
 As base stations (BSs)  equipped with    a large number of antennas become more common in the cellular networks \cite{Larsson&Edfors&Tufvesson&Marzetta:ICM:14}, multicast beamforming can be  judiciously exploited  to support
content multicasting in future wireless applications. In this work, we consider the key problem of group scheduling for downlink multicast transmission. When there are many groups with more users than the available BS antennas in the system, the BS needs to schedule   these groups  over different time slots effectively, in combination with optimized multicast beamforming, to maximize the user throughput.  Furthermore, it is  essential that joint group scheduling and multicast beamforming is scalable with low computational complexity, allowing their application to large-scale wireless systems.

Existing works   on multicast beamforming have mainly focused on the beamforming design at the BS with various  performance objectives or network configurations. The family of multicast beamforming problems are generally nonconvex and NP-hard \cite{Sidiropoulos&etal:TSP2006}.
Thus, finding an effective suboptimal  multicast beamforming solution has been the main challenge. Existing works have developed various approaches to  find   approximate  solutions \cite{Sidiropoulos&etal:TSP2006,Karipidis&etal:TSP2008,Ottersten&etal:TSP14,Xiang&Tao&Wang:IJWC:13},  to improve the beamforming performance \cite{DongLiang:CAMSAP13,Tran&etal:SPL2014,Mehanna&etal:2015,Christopoulos&etal:SPAWC15,EbrahimiDong:Asilomar23,Mohamadi&etal:TSP22}, and to reduce the computational complexity   \cite{Sadeghi&etal:TWC17,Chen&Tao:ITC2017,Yu&Dong:ICASSP18,Ibrahim&etal:TSP2020,Dong&Wang:TSP2020,Zhang&etal:WCL2022,Zhang&etal:TSP2023,Shadi&etal:TSP2022,Mohamadi&etal:TSP22,EbrahimiDong:Asilomar23}. These works typically consider underloaded systems with only a small number of groups of users that can be served simultaneously. None of them consider the group scheduling aspect in optimizing the network performance.     For next-generation massive user access, the BS needs to serve many active groups in the system by  scheduling these groups over multiple time slots.  However, this adds substantial design challenges to the already complicated multicast beamforming problems, as group scheduling is a combinatoric optimization problem.

User scheduling, for the conventional multi-user downlink transmission of  dedicated data, via  unicast beamforming has been   studied in many works \cite{Yoo&Goldsmith:2006JSAC,Zhang&etal:TWC2012Scheduling,Femenias&etal:TCOMM2016,Zhang&etal:TWC2017SumRate,Fuchs&etal:2005,Razaviyayn&etal:TSP2014,Nguyen&etal:Access2017}.
The BS needs to optimally select a subset of users in each time slot in combination of  specific beamforming strategies to maximize  certain network utility objective while ensuring certain fairness among users. Various user selection algorithms  have been proposed
\cite{Yoo&Goldsmith:2006JSAC,Zhang&etal:TWC2012Scheduling,Femenias&etal:TCOMM2016,Zhang&etal:TWC2017SumRate,Fuchs&etal:2005,Dimic&etal:TSP2005,Shen&etal:2006,Chen&etal:TSP2008}.
These algorithms explore the  user channels in the spatial domain to predict the level of interference to each other in order to determine the best set of selected users. However, they cannot be directly applied to group scheduling for multicast beamforming. This is because the existing  approaches typically utilize  user channels as user spatial signatures  to determine the level of separation or correlation among users. Such approaches can be justified by the structures of unicast beamforming, where both the optimal structure and common low-complexity schemes (such as the zero-forcing beamforming strategy \cite{Yoo&Goldsmith:2006JSAC}),  are all functions of user channels that are well understood. However,  the notion of   spatial signature becomes unclear for multicast beamforming towards a group of users.

When both scheduling and multicast beamforming design need to be considered at the BS, computational complexity is the
main issue
for massive multiple-input multiple-output (MIMO) systems
with a relatively large number of groups.   Most existing multicast beamforming algorithms rely on  optimization-based computational techniques, and the optimal structure
of multicast beamforming has only been obtained recently in \cite{Dong&Wang:TSP2020}.  Although this structure has been utilized to improve the beamforming design efficiency
 \cite{Zhang&etal:WCL2022,Zhang&etal:TSP2023,Shadi&etal:TSP2022}, whether it can facilitate group scheduling has not been investigated. An efficient and effective approach for joint group
scheduling and multicast beamforming suitable for practical
implementation is necessary but challenging. Aiming at this goal,  in this work, we explore the structure of multicast beamforming to develop  low-complexity  group scheduling techniques for downlink  multicast transmission to maximize the minimum user throughput.

\subsection{Related Work}
The  literature  on downlink multi-group multicast beamforming has mainly focused on multicast beamforming design at the BS to minimize the transmission power,  maximize the minimum signal-to-interference-and-noise ratio (SINR) or minimum rate, or sum group rate.
Earlier works widely adopted semi-definite relaxation (SDR) \cite{Sidiropoulos&etal:TSP2006,Karipidis&etal:TSP2008,Ottersten&etal:TSP14,Xiang&Tao&Wang:IJWC:13} for the traditional multi-antenna systems.
As the number of antennas grew, successive convex approximation (SCA) \cite{Tran&etal:SPL2014,Mehanna&etal:2015,Christopoulos&etal:SPAWC15} became a more attractive approach for its advantages in both  computation and performance over SDR.
As the wireless systems evolve, more recent research efforts have focused on  providing efficient solutions suitable for
large-scale massive MIMO systems, where different design approaches or optimization techniques were proposed to reduce the computational complexity   \cite{Sadeghi&etal:TWC17,Chen&Tao:ITC2017,Yu&Dong:ICASSP18,Ibrahim&etal:TSP2020,Mohamadi&etal:TSP22}. The optimal multicast beamforming structure was then obtained \cite{Dong&Wang:TSP2020}, and it was  further utilized to develop fast computational algorithms with near-optimal performance for large-scale systems \cite{Zhang&etal:WCL2022,Zhang&etal:TSP2023,Shadi&etal:TSP2022,EbrahimiDong:Asilomar23}.
These works commonly assume that  all groups are served simultaneously. None of them address the problem of group scheduling in networks when the BS needs to serve many active groups over multiple time slots.
To the best of our knowledge, group scheduling with multicast beamforming has not been studied in the literature.

  Different from the fixed user-group association considered in the above works,   several works assume flexible user-group association and have  studied the problem of user grouping, \ie how to assign users into different multicast groups
\cite{Zhou&Tao:2015ICC,Hu&etal:Netw2018,Bandi&etal:2020TWC,Bandi&etal:TWC2022,Tran&Yue:2019Globecom,Yue&Qi:VTC2020,Fuente&etal:2021ICC,Chris&etal:2015TWC} to maximize the multicast beamforming performance.
Message-based user grouping was considered in \cite{Zhou&Tao:2015ICC,Hu&etal:Netw2018,Bandi&etal:2020TWC,Bandi&etal:TWC2022},
where each user can be assigned to one of the groups to receive the message dedicated to that group. In  \cite{Hu&etal:Netw2018,Bandi&etal:2020TWC,Bandi&etal:TWC2022}, admission control was further considered by proposing different optimization methods for joint user selection, user grouping, and multicast beamforming. To address the  issue of performance  deterioration faced by a large multicast group, the works in  \cite{Tran&Yue:2019Globecom,Yue&Qi:VTC2020,Fuente&etal:2021ICC} proposed coding-based user grouping methods to divide users into multiple groups, where
each group adopts a unique modulation and coding scheme that is different from other groups. Heuristic greedy-based algorithms were studied in \cite{Tran&Yue:2019Globecom}, and clustering methods based on user channel spatial correlation were proposed in \cite{Yue&Qi:VTC2020,Fuente&etal:2021ICC}. Finally,  multicast beamforming was utilized for satellite communications to send the coded frames to different groups in \cite{Chris&etal:2015TWC}, where a user grouping
method based on the levels of user channel correlation  was proposed. Note that these works still assume all groups are served simultaneously via multicast beamforming, and the problems addressed are different  from group scheduling over time slots.

For multi-user downlink dedicated data transmission via unicast beamforming,  many existing works have studied user scheduling with specific beamforming strategies
\cite{Yoo&Goldsmith:2006JSAC,Zhang&etal:TWC2012Scheduling,Femenias&etal:TCOMM2016,Zhang&etal:TWC2017SumRate,Fuchs&etal:2005,Razaviyayn&etal:TSP2014,Nguyen&etal:Access2017}.
 As the network throughput can be maximized by  optimally selecting a subset of users in each time slot, various low-complexity greedy-type user selection algorithms were proposed in \cite{Yoo&Goldsmith:2006JSAC,Zhang&etal:TWC2012Scheduling,Femenias&etal:TCOMM2016,Zhang&etal:TWC2017SumRate,Fuchs&etal:2005,Dimic&etal:TSP2005,Shen&etal:2006,Chen&etal:TSP2008}.
In \cite{Yoo&Goldsmith:2006JSAC},  a semi-orthogonal user selection (SUS) method
was proposed to maximize the sum rate of the set of selected users.
It measures the spatial separation of  user channels to form a candidate user set and selects the users with the
largest channel gains from the set.
User selection was extended to both time and frequency scheduling in
\cite{Yoo&Goldsmith:2006JSAC,Zhang&etal:TWC2012Scheduling,Femenias&etal:TCOMM2016,Zhang&etal:TWC2017SumRate,Fuchs&etal:2005,Razaviyayn&etal:TSP2014,Nguyen&etal:Access2017},
where a group of users are selected for each frequency channel and  time slot. The joint optimization problems
of user scheduling and beamforming were formulated and
solved by optimization-based methods in
 \cite{Razaviyayn&etal:TSP2014,Nguyen&etal:Access2017}.
However, both algorithms have high computational complexity.
As mentioned earlier, these user selection and scheduling methods for unicast beamforming cannot be applied to our problem of group scheduling with multicast beamforming.

\subsection{Contribution}
In this paper,  we  address the problem of joint group scheduling and multicast beamforming to maximize the minimum user throughput. We consider fixed user-group association for the multicast groups and for the design constraints on scheduling and the transmit power.  We explore the optimal multicast beamforming structure and use the \emph{group-channel direction} from  the optimal structure  as the spatial
signature of each group for scheduling. The main contributions are summarized as follows:
\begin{mylist}
\item
We propose a three-phase approach to tackle the joint optimization problem
efficiently. Phase 1 utilizes the optimal multicast beamforming structure to obtain the group-channel directions for all groups efficiently, which are then used in Phase 2 to schedule spatially dissimilar groups into the same time slot, followed by  generating the multicast beamformers   in Phase 3 for the scheduled groups via a computationally efficient method.
We observe that this approach provides a computationally efficient solution, whereas the standard alternating optimization approach fails since both the group scheduling and multicast beamforming problems are nonconvex and NP-hard.  The group-channel directions generated in Phase 1   serve as the effective spatial signatures of the groups to be used to measure the inter-group interference in the subsequent scheduling phase.

\item  We propose two low-complexity scheduling algorithms to determine the subset of groups for each time slot and the total number of time slots for Phase 2. The first algorithm is named multi-group multicast scheduling via group spatial
separation (MGMS-GSS).
It  measures the level of spatial separation
among groups and selects spatially dissimilar groups into the same time slot to maintain low interference.
In particular, MGMS-GSS uses a
group-spatial-separation-based (GSS) selection method
to select a subset of groups in each time slot. GSS  uses a  semi-orthogonality
metric to measure the level of spatial separation among groups based on the group-channel directions. It determines a set of semi-orthogonal groups that maximize the minimum rate.
The second scheduling algorithm is named multi-group multicast scheduling via group spatial
correlation (MGMS-GSC).
It uses a design strategy  opposite to MGMS-GSS to  maintain low interference in a subset of groups scheduled in a time slot.
MGMS-GSC first identifies the spatially  correlated groups and then separates them into different time slots. Specifically,  a
group-spatial-correlation-based (GSC)  clustering method is proposed to form clusters of similar groups. GSC is built on a mean-shift-based unsupervised learning technique to capture the similar groups using a spatial correlation metric.
A post-processing procedure is then proposed to assign the spatially correlated groups in the same cluster to different time slots that maximize the minimum user rate within the scheduled groups.
Both MGMS-GSS and MGMS-GSC schedule the subset of groups in each time slot sequentially without extra scheduling delay.

\item  Simulation results show that
both
MGMS-GSS and MGMS-GSC can capture the level of spatial separation  among groups based on the degrees of freedom available to effectively determine the required number of time slots and the set of scheduled groups to improve the minimum user throughput, as compared with  scheduling all groups in a single time slot or one group per time slot. Furthermore, both methods
have low computational complexity in obtaining the
scheduling decision.
Comparing the two,
MGMS-GSS achieves
higher minimum user throughput than MGMS-GSC,
while MGMS-GSC has a  lower computational complexity and is more scalable than MGMS-GSS
as the number of BS antennas increases

\end{mylist}

\subsection{Organization and Notations}
The rest of this paper is organized as follows.
Section~\ref{sec:system_prob}
presents the system model and joint group scheduling and  multicast beamforming problem formulation.
In Section~\ref{sec:joint_algorithm}, we propose a three-phase
design approach.
Sections~\ref{sec:phase1_TS_channel} presents the method for determining the group-channel direction for each group in Phase 1. In Section~\ref{sec:phase2_TS_assign}, we propose our main scheduling algorithms, MGMS-GSS and MGMS-GSC, for Phase 2. The fast multicast beamforming computation for scheduled groups in Phase 3 is presented in Section~\ref{sec:phase3_MB}.
Simulation results are provided in Section~\ref{sec:simulations},
followed by  the conclusion in Section~\ref{sec:conclusion}.

\textit{Notations}:
Hermitian and transpose are denoted as
$(\cdot)^{H}$ and $(\cdot)^{T}$, respectively.
The Euclidean norm of a vector
is denoted as $\|\cdot\|$.
The identity matrix is denoted
as $\Ibf$.
The notation $|z|$ means the absolute value of
scalar $z$, and the notation $|\Zc|$ means
the number of elements in set $\Zc$.

\allowdisplaybreaks
\section{System Model and Problem Formulation}
\label{sec:system_prob}

We consider a downlink multicast transmission scenario, where
the BS equipped with $N$ antennas sends messages to $G$  groups.
We assume that
group $i$ consists of $K_i$ single-antenna users, who receive
a common message from the BS that is independent of the messages to other groups.
Denote the set of group indices by $\Gc\triangleq\{1,\ldots,G\}$,
the set of user indices in group $i$
by $\mathcal{K}_{i} \triangleq \{1,\ldots,K_i\}$, for $i\in\Gc$,
and the total number of users in all groups by $K_{\text{tot}}\triangleq \sum_{i=1}^{G}K_i$.

We consider a time-slotted system where the time slot is indexed by $t \in \{1, 2, \ldots\}$.
Assume that there is a message to be sent to each group. The BS
schedules $G$ groups, possibly over multiple time slots,
and sends their messages via multicast beamforming in each time slot.
We assume each group is scheduled in exactly one time slot for its message transmission,
and multiple groups may be scheduled in the same time slot.
Consider that the BS schedules these $G$ groups in $T$ time slots, where $T \leq G$.
Let $x_{i,t}\in\{0,1\}$ be the scheduling variable,
where
$x_{i,t}=1$ indicates that group $i$ is scheduled in time slot $t$
and $0$ otherwise.
We use $\Gc_{t} \triangleq \{i \,\, | \,\, x_{i,t} = 1, i\in\Gc\}$
to denote the index set of those groups scheduled in time slot $t\in\Tc\triangleq \{1, \ldots, T\}$,
and
 $G_{t} \triangleq |\Gc_{t}|$ is the corresponding number of scheduled groups, with $\sum_{t=1}^{T}G_{t} = G$.

We consider a slow fading scenario,
where   user channels remain unchanged within $T$ time slots.
Let $\hbf_{ik}\in \mathbb{C}^{N\times 1}$ denote the channel vector from the BS to user $k$
in group $i$ in this $T$-time-slot duration. We assume that the BS has the perfect knowledge of $\{\hbf_{ik}\}$.
Let $\wbf_{i}\in \mathbb{C}^{N\times 1}$ denote the multicast beamforming vector for group  $i\in \Gc_t$  that is scheduled in time slot   $t\in \Tc$.
Then,
the received signal at user $k$ in group $i \in \Gc_t $,   $t\in \Tc$, is given by
\begin{align}
y_{ik} = \wbf^{H}_{i}\hbf_{ik}s_{i} + \sum_{j\neq i,j\in\Gc_{t}}\wbf^{H}_{j}\hbf_{ik}s_{j} + n_{ik}, \quad i\in\Gc_t \nn
\end{align}
where
$s_i$ is the symbol intended for group $i$ with $E[|s_{i}|^{2}] = 1$,\footnote{Note that there can be a sequence of symbols transmitted in time slot $t$. Since the transmitted symbols are i.i.d., we ignore the symbol index within a time slot and use $s_i$ to represent one such symbol sent in time slot $t$, which does not cause any ambiguity. The same applies for the received signal $y_{ik}$ and the receiver noise $n_{ik}$.}
and $n_{ik}$ is the user $k$'s receiver additive white Gaussian noise with zero mean and variance $\sigma^{2}$. The  received SINR at user $k$ in group $i\in\Gc_t$
is given by
\begin{align}
\text{SINR}_{ik,t} = \frac{|\wbf^{H}_{i}\hbf_{ik}|^{2}}{\sum_{j\neq i,j\in\Gc_{t}}|\wbf^{H}_{j}\hbf_{ik}|^{2}+\sigma^{2}}, \quad i\in\Gc_t, \label{eq_sinr}
\end{align}and the corresponding achievable rate is\begin{align}
R_{ik,t} = \log_{2}(1+\text{SINR}_{ik,t}), \quad i\in\Gc_t.  \label{eq_rate_SINR}
\end{align}
Since  $G$ groups are scheduled in $T$ time slots, the throughput achieved at each user is then ${R_{ik,t}}/{T}$.

Our goal is to design the  group scheduling for the multicast beamforming to maximize the minimum user throughput among all users in the system,
subject to the total transmit power  and the scheduling constraints.
This overall joint optimization problem is formulated as
\begin{align}
\Pc_{o}:   \max_{T,\{\xbf_t\}_{t=1}^{T},\wbf} &\,  \min_{t\in \Tc}\min_{i\in\Gc_{t},k\in\Kc_{i}} \, \frac{R_{ik,t}}{T}  \nn\\
\text{s.t.} & \ \ x_{i,t}\in \{0, 1\}, \  i\in \Gc, t\in \Tc \label{constraint:binary}\\
&  \ \  \sum_{t=1}^{T}x_{i,t}=1,\  i\in \Gc \label{constraint:group}\\
&  \ \   \sum_{i\in\Gc_{t}}\|\wbf_{i}\|^{2}\leq P, \ t\in \Tc \nn
\end{align}
where
 $\wbf$ $\triangleq$  $[\wbf^H_1,$ $\ldots,$ $\wbf^H_G]^H$ is the concatenated beamforming vectors of all groups,
  $\xbf_{t} \triangleq [x_{1,t},\ldots,x_{G,t}]^T$ is the scheduling decision vector in time slot $t$, and $P$ is the transmit power budget at the BS.
Constraint \eqref{constraint:group}  ensures that each group $i$ is scheduled in exactly one time slot within $T$ time slots.

Problem $\Pc_{o}$ is a mixed-integer programming problem, due to binary scheduling variables. Furthermore, it has a max-min objective, and the rate expression is nonconvex with respect to (w.r.t.) the beamforming vector $\wbf$. As a result, the problem is nonconvex  NP-hard and challenging  to solve.
In the next section, we propose a three-phase approach to compute
a high-quality solution  for problem $\Pc_{o}$.

\section{Three-Phase Optimization Approach}
\label{sec:joint_algorithm}

To make the joint optimization problem $\Pc_o$ more tractable,
we may consider decomposing $\Pc_o$ into two subproblems:
 the scheduling subproblem and the multi-slot multicast beamforming subproblem, which are described as follows:
\begin{mylist}
\item
\emph{Scheduling}: Given the multicast beamforming vector $\wbf$ of all groups, optimizing the scheduling decision $(T,\{\xbf_t\})$ for  $G$ groups  as
\begin{align}
\Pc^{\text{sc}}_{1}(\wbf): & \max_{T,\{\xbf_t\}_{t=1}^{T}} \, \min_{t\in \Tc}\min_{i\in\Gc_{t},k\in\Kc_{i}} \, \frac{R_{ik,t}}{T}  \nn\\
& \,\,\text{s.t.} \,\,  x_{i,t}\in \{0, 1\}, \,\, i\in \Gc, t\in \Tc \nn\\
& \quad\,\,\,\,  \sum_{t=1}^{T}x_{i,t}=1,\,\, i\in \Gc \nn.
\end{align}

\item \emph{Multi-slot multicast beamforming}: Given the scheduling decision $T$  and $\{\xbf_t\}$,
 optimizing the multicast beamforming vector $\wbf$ of  all $G$ groups as
\begin{align}
\Pc^{\text{bf}}_{1}(T, \{\xbf_t\}): & \max_{\wbf} \, \min_{t\in \Tc}\min_{i\in\Gc_{t},k\in\Kc_{i}} \, R_{ik,t}  \nn\\
& \,\,\text{s.t.} \,\, \sum_{i\in\Gc_{t}}\|\wbf_{i}\|^{2}\leq P, \,\, t\in \Tc. \label{constraint_power_multiS}
\end{align}
\end{mylist}

It is clear that the two subproblems are highly intertwined, as $\wbf$ determines how well different groups can be separated  spatially via multicast beamforming, which affects  the scheduling decision ($T, \{\xbf_t\}$), and vice versa.

 Note that one may consider applying the widely-used alternating optimization approach to the above two subproblems
to solve $\wbf$ and $\{\xbf_t\}$ alternatingly.
However, we note that both two subproblems are  nonconvex  NP-hard.
In particular, $\Pc^{\text{sc}}_{1}(\wbf)$ contains
binary scheduling variables and is a max-min optimization problem, and  $\Pc^{\text{bf}}_{1}(T, \{\xbf_t\})$  is a multi-slot
max-min fair (MMF) multicast beamforming problem, which is challenging to solve.\footnote{The single-slot multi-group MMF problem is a difficult problem that has been widely studied in the literature, and the existing algorithms can only guarantee to find stationary points.}
Thus,  applying alternating optimization between   $\Pc^{\text{sc}}_{1}(\wbf)$ and
$\Pc^{\text{bf}}_{1}$$(T,$ $\{\xbf_t\})$
 not only incurs high computational complexity, especially
for large-scale problems, but also may not converge.
 Thus, we need to develop a different approach. In particular,  a good indication of potential inter-group interference is essential at the scheduling stage. However, as multicast beamforming is a complicated problem itself, determining inter-group interference is highly nontrivial.

\begin{algorithm}[t]
\caption{Three-Phase Algorithm for $\Pc_{o}$}
\label{alg:joint}
\begin{algorithmic}[1]
\STATE  {\bf Phase 1: Determining group-channel directions}
\STATE Compute the group-channel direction $\hat{\hbf}_i$ for each group $i\in\Gc$ using \eqref{eq:group_channel}.
\STATE {\bf Phase 2: Scheduling groups}
\STATE Determine the scheduling decision $(T,\{\xbf_t\})$ using $\{\hat{\hbf}_i, i\in\Gc\}$ via MGMS-GSS or MGMS-GSC.
\STATE  {\bf Phase 3: Generating multicast beamformers}
\STATE Solve $\Pc^{\text{bf}}_{1}(T, \{\xbf_t\})$ to determine the multicast beamforming vector $\wbf$.
\RETURN $\{(T,\{\xbf_t\})$, $\wbf\}$
\end{algorithmic}
\end{algorithm}

In this work, aiming at an efficient design, we utilize the characteristics of the optimal multicast beamforming structure and propose a three-phase approach to separate the scheduling and beamforming subproblems to find a solution for $\Pc_o$. The three phases are further described as follows:
\begin{mylist}
\item \emph{Phase 1: Determining group-channel directions.}
Utilizing the optimal multicast beamforming structure, we first determine the \emph{group-channel direction} for each group $i$,  based on the user channels  $\{\hbf_{ik}, k\in\Kc_i\}$ in the group. It    approximately indicates the direction of  beamformer $\wbf_i$ for  group $i$. The group-channel directions will provide the relative degree of spatial separation among the $G$ groups, indicating the potential level of inter-group interference  if any groups are scheduled in the same time slots.

\item \emph{Phase 2: Scheduling groups.} Based on the group-channel directions obtained in Phase 1, we determine the scheduling  decision $(T,\{\xbf_t\})$ for the $G$ groups.
We  propose two low-complexity scheduling schemes, namely
MGMS-GSS and MGMS-GSC. MGMS-GSS uses the notion of semi-orthogonality to assign the groups with mutually semi-orthogonal channel directions
in the same time slot to reduce the inter-group interference.
MGMS-GSC is based on the notion of clustering to first determine  groups with highly correlated group-channel directions. Then,
a post-processing procedure is  performed to assign the groups from the same cluster to the different time slots.

\item \emph{Phase 3: Generating multicast beamformers.} Based on the scheduling decision, we solve the multi-slot
MMF multicast beamforming problem  $\Pc^{\text{bf}}_{1}(T, \{\xbf_t\})$ to determine the beamforming vector $\wbf_i$
for each group $i$.
\end{mylist}

Our proposed three-phase optimization approach for $\Pc_{o}$ is summarized in
Algorithm~\ref{alg:joint}.
In the following sections, we describe the detail of each
phase.

\section{Phase~1: Determining Group-Channel Directions}
\label{sec:phase1_TS_channel}

We first determine the group-channel direction for each group $i\in\Gc$ based on all the user channels  $\{\hbf_{ik}$'s,  $k\in\Kc_{i}\}$ in the group.
The notion of the group-channel direction was first introduced in \cite{Dong&Wang:TSP2020}, where the optimal
multicast beamforming structure has been obtained for   the BS serving multiple groups simultaneously in the same time slot.
Specifically, consider  $G_t$ groups in $\Gc_t$ in time slot $t$.  It is shown in \cite{Dong&Wang:TSP2020} that
the optimal multicast beamforming solution to the following MMF problem for $\Gc_t$
\begin{align}
\Sc^{t}_{o}: \max_{\{\wbf_i, i\in\Gc_{t}\}}&\ \min_{i\in\Gc_{t},k\in\Kc_i}  \text{SINR}_{ik,t} \nn \\
& \text{s.t.} \ \sum_{i\in\Gc_{t}}\|\wbf_{i}\|^{2}\leq P. \nn
\end{align}
 has a weighted MMSE beamforming structure given by
\begin{align}
\wbf_{i} = \Rbf^{-1}\Hbf_{i}\abf_{i} ,\quad i\in\Gc_{t}
\label{optimal_structure_BF}
\end{align}
where $\Rbf$ is the  noise-plus-weighted-channel-covariance matrix, which is  a function of $\hbf_{ik}$'s of all users in these $G_t$ groups   and the transmit power to noise ratio $P/\sigma^2$,   $\Hbf_{i} \triangleq [\hbf_{i1}, \ldots, \hbf_{iK_{i}}]$ is the channel matrix for group $i$, and
$\abf_{i}\in \mathbb{C}^{K_i\times 1}$ is the optimal weight vector for group $i$. Note that based on  \eqref{eq_rate_SINR}, the max-min SINR objective in $\Sc_o^t$ is equivalent to  the max-min user rate optimization.

For the optimal solution in \eqref{optimal_structure_BF}, the term $\Hbf_{i}\abf_{i}$ forms the group-channel direction, defined by
\begin{align}
\hat{\hbf}_i \triangleq \Hbf_{i}\abf_{i}= \sum_{k=1}^{K_{i}} a_{ik}\hbf_{ik}. \label{eq:group_channel}
\end{align}
It is a weighted sum of all user channels in group $i$ with weight $a_{ik}$ being the $k$-th element in $\abf_{i}$, indicating
the relative significance of user channel $\hbf_{ik}$
in $\hat{\hbf}_i$. Thus,  we have $\wbf_i = \Rbf^{-1}\hat{\hbf}_i$, where the group-channel direction  $\hat{\hbf}_i$ indicates the direction that the optimal multicast beamforming vector $\wbf_i$ for group $i$ is beamforming to.

Moreover, we note that the set of group-channel directions  $\{\hat{\hbf}_i\}_{i\in\Gc_t}$ also indicate the degree of spatial separation among these $G_t$ groups, reflecting the potential level of inter-group interference.
For group scheduling, our aim is to control the inter-group interference at a minimum level in each time slot. The set of $\hat{\hbf}_i$'s  provides an effective measure of the level of inter-group interference. Thus, we propose to use this group-channel direction as a spatial signature to represent each group to facilitate the group scheduling in Phase~2.

However, for the scheduling purpose, determining the group-channel direction is not straightforward. In particular,  for the optimal $\wbf_i$ in \eqref{optimal_structure_BF},  weight vector $\abf_i$ for $\hat{\hbf}_i$ need to be optimized jointly among all the groups that are scheduled in the same time slot  \cite{Dong&Wang:TSP2020}, which is only known after the scheduling decision is made.
Therefore, the actual group-channel direction  cannot be obtained  in this phase a priori.
Nonetheless, since our goal   is to schedule groups base on the level of spatial separation  among groups to minimize  inter-group interference,
 we
propose to obtain the  group-channel direction
treating each group as the only group in the multicast system
without considering other groups.

\subsection{Single-Group-Based Group-Channel Direction }\label{susbection:single_group_MMF}
Following the above discussion,
we now determine $\hhbf_i$ for each group $i\in \Gc$ without considering the other groups.
Equivalently, we consider the following single-group MMF problem w.r.t. $\wbf_i$:
\begin{align}
\Sc_{1,i}:   \;\max_{\wbf_{i}} \min_{k\in\Kc_{i}} \;\;& |\wbf_{i}^{H}\hbf_{ik}|^{2} \nonumber\\
   \text{s.t.} \;\; & \|\wbf_{i}\|^2\leq P.  \nn
\end{align}
Since we only consider  group $i$ in $\Sc_{1,i}$, i the optimal solution structure in \eqref{optimal_structure_BF},  sthe  noise-plus-weighted-channel-covariance matrix $\Rbf$  in \eqref{optimal_structure_BF} only contains  $\hbf_{ik}$'s in group $i$. Thus, we use $\widetilde{\Rbf}_i$ to represent $\Rbf$ in this case to indicate its dependency on group $i$ only. Following this, we  convert $\Sc_{1,i}$  into a weight optimization problem  w.r.t. $\abf_i$, given by
\begin{align}
\Sc_{2,i}:   \;\max_{\abf_i} \min_{k\in\Kc_i} \;\;& |\abf_i^H\Hbf^H_i\widetilde{\Rbf}_i^{-1}\hbf_{ik}|^2 \nonumber\\
   \text{s.t.} \;\; & \|\widetilde{\Rbf}_i^{-1}\Hbf_i\abf_i\|^2\leq P.  \nn
\end{align}
Once  $\abf_i$ is obtained, we can then determine $\hhbf_i$ by \eqref{eq:group_channel}.

Note that for massive MIMO\ systems with $N\gg 1$, the size of the weight optimization problem  for $\abf_i$ is significantly smaller than $\Sc_{1,i}$ ($K_i \ll N$). However, $\Sc_{2,i}$ is  still a nonconvex and NP-hard problem, and we need to solve $G$ such problems for all $i\in\Gc$. Therefore, it is important that we can compute $\hat{\hbf}_i$, $i\in\Gc$, efficiently in this phase. In a recent work   \cite{Zhang&etal:WCL2022},  a  fast first-order algorithm based on  the optimal  structure in \eqref{optimal_structure_BF} and the  projected subgradient algorithm (PSA)
has been proposed for the  MMF problem $\Sc^{t}_{o}$. It provides a near-optimal performance with significantly lower computational complexity than the other existing algorithms.  We can directly employ this algorithm to solve $\Sc_{1,i}$ efficiently.

In particular, the algorithm in  \cite{Zhang&etal:WCL2022} uses an approximate closed-form expression for  $\widetilde{\Rbf}_i$  for fast computation. Express each channel  as $\hbf_{ik}=\sqrt{\beta_{ik}}\gbf_{ik}$, where $\beta_{ik}$ is the channel variance,  and $\gbf_{ik}$ is the normalized channel vector with unit variance and  i.i.d. zero mean elements representing the small-scale fading. The  approximate expression is given by
\begin{align}
\widetilde{\Rbf}_i  \approx {\bf{I}}+\frac{P\tilde{\beta}_i}{\sigma^{2}K_{i}}\sum_{k=1}^{K_{i}}\gbf_{ik}\gbf_{ik}^{H}
\label{eq:asymp_R_single}
\end{align}
where $\tilde{\beta}_i \triangleq 1/(\frac{1}{K_i}\sum_{k=1}^{K_i}\frac{1}{\beta_{ik}})$
is the harmonic mean of the channel variances of all users in group $i$.
 Using $\widetilde{\Rbf}_i$ in \eqref{eq:asymp_R_single}, we can compute $\abf_i$ in
$\Sc_{2,i}$ using the PSA-based iterative algorithm in \cite{Zhang&etal:WCL2022},
which uses only closed-form updates   and is guaranteed to find a near-stationary point of $\Sc_{2,i}$.
\section{Phase 2: Scheduling Groups }
\label{sec:phase2_TS_assign}

In this phase, we  propose two low-complexity algorithms to determine the scheduling decision ($T, \{\xbf_t\}$).
Since the group-channel direction $\hhbf_{i}$ characterizes the spatial direction of group $i$ for beamforming, the two algorithms use   $\{\hhbf_{i}\}$ to determine which groups can be scheduled in the same time slots. However, they adopt two opposite design strategies
for maintaining low interference in each time slot.

\subsection{Multi-Group Multicast Scheduling via Group Spatial Separation }
\label{sec:SIE}

MGMS-GSS measures the level of spatial separation among  groups and selects dissimilar groups into the same time slot in a sequential manner, \ie $\xbf_1,\xbf_2,\ldots$, are formed sequentially. The total number of time slots $T$ required  is  determined automatically at the end when all the $G$ groups are scheduled. Such sequential scheduling can be implemented  per time slot in real-time at the BS, minimizing the scheduling delay for the $G$ groups.

Let $\Uc_t$ denote the index set of the unscheduled groups after time slot $t-1$.
 Starting at time slot $t=1$, with the initial set   $\Uc_1=\Gc$,\footnote{As indicated in $\Pc_o$, we note that the time slot index $t$ is w.r.t. the $T$-slot  scheduling epoch of the $G$ groups, \ie $t=1,\ldots,T$. } MGMS-GSS measure the level of spatial separation among the  groups in  $\Uc_t$ to determine\ the  groups    $\Gc_{t}$ to be scheduled in the current time slot $t$. Note that $\Gc_t$ contains the same information as $\xbf_t$, and thus, we use them interchangeably.  We first introduce the definition of semi-orthogonality \cite{Yoo&Goldsmith:2006JSAC} below.

\begin{definition}[Semi-orthogonality]\label{def:SO}
Given $\zbf,\zbf' \in \mathbb{C}^{N\times 1}$ and a positive constant $\alpha\in(0,1]$,
vectors $\zbf$ and $\zbf'$ are said to be semi-orthogonal to each other  if
\begin{align}\label{eq:SO}
\frac{|\zbf^{H}\zbf'|}{\|\zbf\| \|\zbf'\|}<\alpha.
\end{align}
\end{definition}
We now describe a group spatial
separation (GSS) method for the group selection.

\subsubsection{Semi-orthogonal group selection} GSS uses an  iterative procedure for group selection.
In each iteration, it uses the group-channel directions $\{\hhbf_{i}\}$ to  measure semi-orthogonality among the unselected groups to form a set of semi-orthogonal groups and then select one into $\Gc_t$.
The  procedure is  repeated until no more groups can be further selected.
There are two main steps at each iteration $n$: i)  Group selection; ii) Candidate group set update. We describe each step below.

\emph{i) Group selection:}
Let $\Gamma^{(n)}$ denote the set of the candidate  groups  at iteration $n$ for selection, with initial $\Gamma^{(1)} = \Uc_{t}$.
 How to determine  $\Gamma^{(n)}$  will be discussed in the next step. Note that  at the beginning of iteration $n$,  $\Gc_t$ contains the  selected groups up to iteration $n-1$, and $\Gc_t \cap \Gamma^{(n)}=\emptyset$.
Let $i_n^{\star}$ denote the index of the  group selected at iteration $n$. Based on the max-min throughput objective of $\Pc_o$ for scheduling, we select a group  $i_n^{\star}  \in \Gamma^{(n)}$ to maximize the minimum achievable rate among the scheduled groups in the current time slot $t$. Specifically, assume $i\in\Gamma^{(n)}$ is selected, and let $\widetilde{\Gc}_t^i \triangleq \Gc_t \cup \{i\}$.
The max-min rate for $\widetilde{\Gc}_t^i$ is obtained by optimizing the multicast beamforming vectors $\{\wbf_j, j\in \widetilde{\Gc}_t^i\}$ to  maximize the minimum SINR among the groups in $\widetilde{\Gc}_t^i$, \ie the MMF problem similar to  $\Sc_o^t$ in Section~\ref{sec:phase1_TS_channel}, given by
\begin{align}
\widetilde{\Sc}_{i}^t: \max_{\{\wbf_j: j\in\widetilde{\Gc}_t^i\}}&\min_{j\in\widetilde{\Gc}_t^i,k\in\Kc_j} \ \text{SINR}_{jk,t} \nn\\
&\text{s.t.} \  \sum_{j\in\widetilde{\Gc}_t^i}\|\wbf_{j}\|^{2}\leq P. \nn
\end{align}
We solve the above problem for each $i\in \Gamma^{(n)}$.
Let $\gamma_{\min,i}^\star$ be the corresponding minimum SINR maximized in  $\widetilde{\Sc}_{i}^t$, $i\in \Gamma^{(n)}$. Then, the selected group is given by
\begin{align}\label{eq_modi_SGS_selec}
i_n^{\star} =  \underset{i\in\Gamma^{(n)}}{\arg\max}\,\, \gamma_{\min,i}^\star.
\end{align}
Following this, we update $\Gc_t$ as $\Gc_t \leftarrow \Gc_t \cup \{i_n^{\star}\}$.

The above approach requires solving $\widetilde{\Sc}_{i}^t$      for each $i\in \Gamma^{(n)}$ at each iteration $n$. Thus, it is essential that  the solution to  $\widetilde{\Sc}_{i}^t$ can be obtained efficiently. The optimal solution structure of  $\wbf_j$ for the MMF problem  $\widetilde{\Sc}_{i}^t$ is given in \eqref{optimal_structure_BF}. However, we still need to determine parameters in $\Rbf$ and $\abf_i$ via numerical algorithms. Fortunately,  the asymptotic expression of the optimal $\wbf_j$ as $N\to \infty$  is obtained in closed-form \cite{Dong&Wang:TSP2020}. Since  our main purpose  at this stage is to select a group, we can use this closed-form expression as an approximate solution for $\wbf_j$  to  select a group selection with low complexity.

Specifically, the asymptotic beamforming solution for group $j\in \widetilde{\Gc}_{i}^t$ is given by \cite{Dong&Wang:TSP2020}
\begin{align}
\wbf_{j} = c_{j}\bar{\Rbf}^{-1}\Hbf_{j}\qbf_{j}
\label{eq:SGS_approx_BF}
\end{align}
where $\qbf_{j}\triangleq [1/\beta_{j1}, \ldots, 1/\beta_{jK_{j}}]^{T}$ with $\beta_{jk}$ being the channel variance of each user defined below \eqref{eq:asymp_R_single},
and $\bar{\Rbf}$ is given by the following closed-form expression, which is a generated version of  \eqref{eq:asymp_R_single} for the single-group case:
\begin{align}
\bar{\Rbf} = {\bf{I}}+\frac{P\bar{\beta}}{ \sigma^{2}\sum_{i\in\widetilde{\Gc}_t^i}K_{i}}\sum_{i\in\widetilde{\Gc}_t^i}\sum_{k=1}^{K_{i}}\gbf_{ik}\gbf_{ik}^{H}\nn
\end{align}
where $\bar{\beta}$ is the harmonic mean of channel variances of users in $\widetilde{\Gc}_{i}^t$  and $c_j$ is the scaling factor:
\begin{align*}
\bar{\beta} \triangleq\! \frac{\sum_{j\in\widetilde{\Gc}_t^i}K_{j}}{ \sum_{j\in\widetilde{\Gc}_t^i}\sum_{k=1}^{K_{j}}\frac{1}{\beta_{jk}}},  c^2_j \triangleq \!\frac{P\sum_{k=1}^{K_j}\frac{1}{\beta_{jk}}}{\sum_{j\in \widetilde{\Gc}_t^i}\sum_{k=1}^{K_j}\!\frac{1}{\beta_{jk}}\|\bar{\Rbf}^{-1}\!\Hbf_j\qbf_j\|^2}.
\end{align*}
Using  \eqref{eq:SGS_approx_BF} as  the approximate solution, we can directly  evaluate $\text{SINR}_{jk,t}$ for each user $k$ in group $j \in \widetilde{\Gc}_t^i$,  determine $\gamma_{\min,i}^\star = \min_{j\in\widetilde{\Gc}_t^i,k\in\Kc_j} \text{SINR}_{jk,t}$, for $i\in\Gamma^{(n)}$, and obtain $i_n^\star$ by \eqref{eq_modi_SGS_selec} .

\indent \emph{ii) Candidate group set update:}
To update the set of candidate groups $\Gamma^{(n+1)}$ for the next iteration $n+1$, we do not just simply remove $i_n^\star$ from $\Gamma^{(n)}$. We also need to pick the groups that are semi-orthogonal to the already selected groups in $\Gc_t$. This is to ensure that the selected groups are always semi-orthogonal to each other to limit the inter-group interference in order to maximize the minimum achievable rate for scheduled groups in the current time slot.

First, based on  $\{\hhbf_i\}$ of the selection groups, we  construct a set of mutually orthogonal vectors over iterations using the Gram-Schmidt procedure. Let $\fbf_1,\ldots,\fbf_{n-1} \in \mathbb{C}^{N\times 1}$ denote the Gram-Schmidt orthonormal vectors formed at  iterations  $1,\ldots,n-1$, where $\fbf_i^H\fbf_j = 0$, $\forall 1\le i,j\le n-1, i\neq j$, and $\|\fbf_i\|=1$, $\forall i$. Based on  $\hhbf_{i_n^\star}$ of the selected group $i_n^\star$,
we form the Gram-Schmidt  vector $\fbf_{n}$ at iteration $n$ as
 \begin{align}
 \fbf_n =  \hhbf_{i_n^\star} - \sum_{j=1}^{n-1}(\fbf^H_{j}\hhbf_{i_n^\star})\fbf_j; \quad \fbf_n \leftarrow \frac{\fbf_n}{\|\fbf_n\|}.
 \label{eq:GS}
 \end{align}
Note that  $\fbf_n$ represents the component of $\hhbf_{i_n^\star}$ that is orthogonal to the subspace spanned by $\{\fbf_1,\ldots,\fbf_{n-1}\}$. By this procedure, we have the set of orthonormal vectors  at iteration $n$ as $\{\fbf_1,\ldots,\fbf_{n}\}$. It reflects the  subspace spanned by the current selected groups in $\Gc_t$.

\begin{algorithm}[t]
\caption{The GSS Method for Determining $\Gc_{t}$}
\label{alg:SGS}
\begin{algorithmic}[1]
\STATE \textbf{Initialization:} Set threshold $\alpha$. Set $n=1$. Set $\Gamma^{(1)} = \Uc_{t}$, $\Gc_t = \emptyset$.
\WHILE {$\Gamma^{(n)}\neq\emptyset$}
\STATE // \textit{Step i): Group selection}
\STATE For each $i\in \Gamma^{(n)}$, compute $\gamma_{\min,i}^\star$ based on $\{\wbf_j: j\in\widetilde{\Gc}_t^i\}$ in \eqref{eq:SGS_approx_BF}.
\STATE Obtain $i_n^{\star}$ by \eqref{eq_modi_SGS_selec}.
 Update $\Gc_t \leftarrow \Gc_t \cup \{i_n^{\star}\}$.
\STATE // \textit{Step ii): Candidate group set update}
       \STATE Compute ${\fbf}_n$ by \eqref{eq:GS} using  $\hbf_{i_n^{\star}}$ and $\{\fbf_1,\ldots,\fbf_{n-1}\}.$
\STATE Update  $\Gamma^{(n+1)}$  by \eqref{eq:SGS_update}.
\STATE Set $n \leftarrow n+1$.
\ENDWHILE
\RETURN \!\!$\Gc_t$
\end{algorithmic}
\end{algorithm}

Next, using the newly added Gram-Schmidt vector $\fbf_{n}$, we determine the set of candidate groups $\Gamma^{(n+1)}$ from $\Gamma^{(n)}$ for the next iteration as
\begin{align}
\Gamma^{(n+1)} = \bigg\{i:  \frac{|\hhbf^{H}_{i} \fbf_{n}|}{\|\hhbf_{i}\|}<\alpha, \ i\in\Gamma^{(n)}, i\neq i_n^{\star}  \bigg\}
\label{eq:SGS_update}
\end{align}
where $\alpha\in(0,1]$ is the  threshold for semi-orthogonality by Definition~\ref{def:SO}.
Note   from \eqref{eq:SGS_update} that at  iteration     $n$, only those groups in $\Gamma^{(n)}$ with $\hat{\hbf}_i$'s that are semi-orthogonal to $\fbf_n$ will be included in the next iteration $n+1$ for consideration. Thus, by this procedure, we see that  at the start of iteration $n+1$, the set of candidate groups $\Gamma^{(n+1)}$ are semi-orthogonal to the existing selected groups in $\Gc_t$ in terms of $\hat{\hbf}_i$.

The proposed GSS repeats Steps i)-ii) to iteratively update $\Gc_t$ until $\Gamma^{(n)}$ is empty, \ie no more unselected groups satisfy the semi-orthogonality condition. We summarize the proposed GSS in Algorithm~\ref{alg:SGS}.

In summary, GSS is a fast greedy-type iterative method for group selection. At each iteration $n$,  GSS first uses Step i) to select a  group into $\Gc_t$  from $\Gamma^{(n)}$, which contains the unselected groups that are semi-orthogonal to  $\Gc_t$. Then, it uses Step ii) to update the set of orthonormal vectors $\{\fbf_1,\ldots,\fbf_n\}$  based on the selected groups and form the next candidate groups that are semi-orthogonal to the  selected groups.
By this iterative procedure, the selected groups in $\Gc_t$ are semi-orthogonal to each other.
As a result, we effectively limit the inter-group interference  and maximize the minimum user rate in the current time slot.

\subsubsection{Scheduling selected groups} For each time slot $t$,
the proposed MGMS-GSS employs the GSS procedure above to obtain $\Gc_t$ (\ie $\xbf_t$), and schedules all selected groups in $\Gc_t$ for transmission. The unselected group set is then updated for the next time slot:
$ \Uc_{t+1} = \Uc_{t}\backslash \Gc_{t}$. The above procedure continues until $\Uc_{t}=\emptyset$, \ie all groups have been scheduled. Then,  the total number of time slots\ $T$  is also determined.

We summarize the proposed MGMS-GSS in Algorithm~\ref{alg:MGMS_general} and have the following remarks.
\begin{algorithm}[t]
\caption{The MGMS-GSS Algorithm for ($T, \{\xbf_t\}$)}
\label{alg:MGMS_general}
\begin{algorithmic}[1]
\STATE \textbf{Initialization:} Set $\Uc_1=\Gc$, $t = 1$.
\WHILE {$\Uc_{t}\neq\emptyset$}
       \STATE Obtain $\Gc_{t}$ and $\xbf_t$ by Algorithm~\ref{alg:SGS}.
       \STATE Update $\Uc_{t+1} = \Uc_{t}\backslash \Gc_{t}$.
       \STATE Set $t \leftarrow t+1$.
    \ENDWHILE
\STATE Set $T = t-1$.
\RETURN \!\!($T, \{\xbf_t\}$)
\end{algorithmic}
\end{algorithm}

\begin{remark}
For MGMS-GSS   sequentially scheduling the $G$ groups, after some time slots, if none of the remaining unscheduled groups are semi-orthogonal to each other, only one group will be selected in $\Gc_t$ based on the GSS procedure. In this case, these remaining groups will be scheduled one at each time slot.
\end{remark}

\begin{remark}
The proposed MGMS-GSS  can be implemented  per time slot in real-time without the need to wait for the scheduling decision of all the $G$ groups over $T$ time slots to be determined.  Thus, it minimizes any scheduling delay at the BS among these $G$ groups. Furthermore, MGMS-GSS is a simple low-complexity algorithm that only involves closed-form computations or evaluation. Thus, real-time scheduling decision can be computed fast at each time slot.
\end{remark}
\begin{remark}
We point out that  semi-orthogonality has first been considered for  user selection in a  multi-user MIMO system in \cite{Yoo&Goldsmith:2006JSAC}, where  the SUS method has been proposed to select users from a user set to maximize the downlink sum-rate. Although both methods are based on semi-orthogonality, some detail of the design strategy in our GSS procedure is different from that in \cite{Yoo&Goldsmith:2006JSAC}: SUS uses individual user channels for user selection, and among the candidate users, the user with the largest channel gain is selected at each user selection iteration. In contrast, our GSS is based on the
group-channel direction of each group and selects a group
that directly maximizes the minimum SINR  among the selected groups by $\widetilde{\Sc}_{i}^t$ and \eqref{eq_modi_SGS_selec}.
Moreover, \cite{Yoo&Goldsmith:2006JSAC} only concerns about the user selection problem in a given time slot, while our MGMS-GSS is a scheduling algorithm of all $G$ groups over multiple time slots.
\end{remark}

\subsection{Multi-Group Multicast Scheduling via Group Spatial Correlation }
\label{sec:HIE}

In contrast to MGMS\_GSS, we now present the second algorithm, MGMS-GSC, uses the clustering idea to first  form clusters, each containing spatially correlated
groups and then separates  these similar groups in the same cluster into different time slots to avoid strong interference to each other.

 We use the clustering technique that uses a similarity metric to find the spatially correlated groups.  In particular, MGMS-GSC is built on the
MS method \cite{Cheng:1995}, a popular unsupervised learning technique that captures the similarity among data points to form clusters.
After forming multiple sets of spatially-correlated groups,
we process these sets to sequentially determine the scheduling decisions $\xbf_1, \xbf_2, \ldots$,
and the total number of time slots $T$, using the max-min user rate objective.

\subsubsection{Preliminaries of mean shift method}
MS is a mode-seeking iterative method to find local maxima in data distribution of a dataset and form data clusters. It determines both the number of clusters and cluster members.
Let $\Yc \triangleq \{\ybf_{i}: \ybf_{i}\in\mathbb{C}^{N\times 1}\}$
denote the dataset (or feature space)
containing the data points $\ybf_{i}$'s.
Let $\cbf$ be the centroid for a cluster based on $\Yc$.
The cluster contains all the data points $\ybf_{i}$'s in $\Yc$ that are within the Euclidean distance $\tau$ from  centroid $\cbf$:
\begin{align}
\|\ybf_{i} - \cbf\|<\tau \label{eq_clustering}
\end{align}
where $\tau>0$ is the similarity threshold affecting the cluster size.
MS obtains centroid $\cbf$ via seeking a local maximum in the
underlying density function of $\Yc$.
The density function of $\Yc$ is estimated by using the
kernel density estimation scheme \cite{Comaniciu&Meer:2002}.
In particular, a kernel $H(\cdot)$ is given by
$H(\ybf) = \mu h(\|\ybf\|^{2})$ for $\ybf\in\Yc$,
where $h(\cdot)$ is the corresponding kernel profile,
and $\mu$ is the normalization factor such that $H(\ybf)$ integrates to 1.\footnote{
The Gaussian kernel is commonly used for $H(\ybf)$ with a profile
given by
$h(\|\ybf\|^{2}) = \exp{(-\|\ybf\|^{2}/2)}$.}
The kernel density estimator (KDE) with kernel $H(\ybf)$
on set $\Yc$ is given by
\begin{align}
\psi(\ybf) = \frac{\mu}{G\tau^{N}}\sum_{i=1}^{G}h
\left(\left\|\frac{\ybf_{i} - \ybf}{\tau}\right\|^{2}\right). \nn
\end{align}
Based on $\psi(\cdot)$, the MS updating procedure
is carried out using the gradient ascent method for finding a local maximum of the KDE function.
In particular, the update for centroid $\cbf^{(l)}$ at iteration
$l$, for $l=1,2,\ldots$, is given by \cite{Comaniciu&Meer:2002}
\begin{align}
\cbf^{(l+1)} = \displaystyle\frac{\displaystyle \sum_{i=1}^{G}
\ybf_{i}h
\left(\left\|\frac{\ybf_{i} - \cbf^{(l)}}
{\tau}\right\|^{2}\right)}{\displaystyle\sum_{i=1}^{G}
h\left(\left\|\frac{\ybf_{i} - \cbf^{(l)}}{\tau}\right\|^{2}\right)}.
\label{eq:ms_centroid_update}
\end{align}
The centroid and the cluster are iteratively updated using the above MS procedure until convergence. This procedure is guaranteed to converge to a local maximum of $\psi(\cdot)$, if the profile
$h(\cdot)$ is convex and monotonically decreasing \cite{Comaniciu&Meer:2002}.

\subsubsection{Group-spatial-correlation-based clustering method}
Based on the MS method, we now propose a  GSC clustering method for the $G$ groups.
It uses the group-channel directions $\{\hat{\hbf}_i\}$ to measure the level of spatial
correlation among the groups and forms multiple clusters, each containing spatially correlated groups.
Specifically, we consider a feature space spanned by the normalized group-channel directions,
given by
\begin{align}
\Yc = \left\{\ybf_{i}: \ybf_{i} \triangleq \frac{\hhbf_{i}}{\|\hhbf_{i}\|}e^{-j\angle{\hat{h}_{1i}}},\, \forall i\in\Gc\right\} \label{eq_data_set}
\end{align}
where $\angle{\hat{h}_{1i}}$ denotes the phase of the first element in vector $\hhbf_{i}$.
Note that each data point $\ybf_{i}$ in $\Yc$ is phase-adjusted such that its first element is phase-aligned to $0$ degree.
This is to guarantee that in the centroid update in \eqref{eq:ms_centroid_update},
all $\ybf_{i}$'s are properly phase-aligned for computing the weighted sum.

The GSC method sequentially generates the clusters using
the MS procedure given in \eqref{eq:ms_centroid_update}.
 In particular, let $R$ denote the number
of clusters that GSC generates in total, and let
$\cbf_r$ be the centroid of the $r$-th cluster.
Denote the set of $\ybf_i$'s in cluster $r$ by
\begin{align}\label{eq_GSC_Yr}
\Yc_{r} = \{\ybf_{i}: \|\ybf_{i} - \cbf_r\|<\tau, \,\,
\forall\ybf_{i}\in\Yc\}.
\end{align}
We employ MS to sequentially obtain clusters
$\Yc_{1}, \Yc_{2}, \ldots$. The number of clusters $R$ formed by the $G$ groups
is automatically determined at the end of the MS procedure.
Let $\Qc_r$ denote the set of remaining $\ybf_i$'s
that are not yet selected by $\Yc_1,\ldots,\Yc_{r-1}$,
and we initialize $\Qc_{1}=\Yc$.
To form cluster $r$ from $\Qc_r$, we
initialize the centroid for cluster $\Yc_r$ as
$\cbf_r^{(1)} \in \Qc_{r}$ and iteratively update the centroid
$\cbf_r$ by \eqref{eq:ms_centroid_update}.
To further simplify the computation,
we adopt a truncated Gaussian kernel profile for the KDE
$\psi(\ybf)$ \cite{Comaniciu&Meer:2002}, given by
\begin{align}
h(\|\ybf\|^{2})\triangleq
\begin{cases}
\exp{(-\|\ybf\|^{2}/2)}         & \text{if} \,\,\, \|\ybf\|<1\text{,} \nn\\
0           & \text{otherwise}\text{.}\nn
\end{cases}
\end{align}
The centroid update $\cbf^{(l+1)}_r$ at iteration $l$ is then given by
\begin{align}
\cbf^{(l+1)}_r \!=\!
\frac{\displaystyle\sum_{\ybf_i\in\Yc_{r}}\!\!\ybf_{i}
\exp{\!\bigg(\!\!-\!\frac{\displaystyle\|\ybf_{i} - \cbf^{(l)}_r\|^2}{\displaystyle2\tau^2}
\!\bigg)}}{\displaystyle\sum_{\ybf_i\in\Yc_{r}}\!\!
\exp{\!\bigg(\!\!-\!\frac{\displaystyle\|\ybf_{i} - \cbf^{(l)}_r\|^2}{\displaystyle2\tau^2}
\!\bigg)}}; \;\; \cbf^{(l+1)}_r \!\leftarrow\!
\frac{\cbf^{(l+1)}_r}{\|\cbf^{(l+1)}_r\|}. \label{eq:ms_centroid_update_truncat}
\end{align}
After the MS procedure converges,
we have $\Yc_r$ as the cluster $r$, and we update set $\Qc_{r+1}$ by
\begin{align}
\Qc_{r+1} = \Qc_{r}\backslash\Yc_{r}. \nn
\end{align}
This sequential clustering procedure continues until
$\Qc_{r+1}=\emptyset$, for some $r$, and we set $R=r$.
We summarize the proposed GSC
in Algorithm~\ref{alg:GSC}.

\begin{algorithm}[t]
\caption{The GSC Method for Determining ($R, \{\Yc_{r}\}$)}
\label{alg:GSC}
\begin{algorithmic}[1]
\STATE \textbf{Initialization:} Set threshold  $\tau$.
Set $\Qc_{1}=\Yc$, $r=1$.
\WHILE {$\Qc_{r}\neq\emptyset$}
       \STATE \textbf{Initialization:} Set
       $\cbf_r^{(1)} \in \Qc_{r}$, $l=1$.
       \REPEAT
       \STATE Compute $\Yc_{r} = \{\ybf_{i}: \|\ybf_{i} - \cbf^{(l)}_r\|<\tau,\,\,\forall\ybf_{i}\in\Yc\}$.
       \STATE Update $\cbf^{(l+1)}_r$ via \eqref{eq:ms_centroid_update_truncat}.
       \STATE Set $l \leftarrow l+1$.
       \UNTIL{convergence}
       \STATE Update $\Qc_{r+1} = \Qc_{r}\backslash\Yc_{r}$.
       \STATE Set $r\leftarrow r+1$.
    \ENDWHILE
\STATE Set $R = r-1$.
\RETURN \!\!($R, \{\Yc_{r}\}$)
\end{algorithmic}
\end{algorithm}

Based on the clustering metric in \eqref{eq_clustering}, each cluster contains groups with their
$\hhbf_i$'s being correlated at a relatively high level.
They will cause more severe
interference to each other and need to be separated into
different time slots.
Next, we use a post-processing procedure to
perform the group scheduling from the $R$ clusters.

\subsubsection{Post-processing procedure}
In this final step, we assign groups from $R$ clusters into a time slot, one from each cluster, to keep a low interference level among the groups in the same time slot. Let $r_{\text{max}}$ be the index of the largest cluster among all $R$ clusters,
and  $G_{\text{max}}\triangleq|\Yc_{r_{\text{max}}}|\le G$ indicates the largest cluster size.
Then, we assign $G$ groups into $G_{\text{max}}$ time slots.  The groups in each given time slot are from different clusters.

In particular, we schedule the groups over time slots sequentially in the order of
$\xbf_1, \ldots, \xbf_{G_{\text{max}}}$.
Let $\Ic_{r}$ be the index set of the groups in $\Yc_r$.
For time slot $t$, we first randomly select a group from
cluster $r_{\text{max}}$,
\ie
$i_{t}\in\Ic_{r_{\text{max}}}$,
and assign it into set $\Gc_t$.
Cluster $r_{\text{max}}$ is updated via
$\Ic_{r_{\text{max}}}\backslash \{i_{t}\}$.
Next, for each of the rest nonempty clusters, \ie $r\neq r_{\text{max}}$, $\Yc_r\neq \emptyset$, we select a group $i^{\star}_r$ from cluster $r$ to maximize the minimum  SINR
(or rate) among the scheduled groups
$\widetilde{\Gc}^i_{t} = \Gc_t \cup \{i\}$:
\begin{align}
i^{\star}_r =  \arg\underset{i\in\Ic_{r}}{\max}\,
\min_{j\in\widetilde{\Gc}^i_{t},k\in\Kc_j}
\frac{|\wbf^{H}_{j}\hbf_{jk}|^{2}}
{\sum_{m\neq j,m\in\widetilde{\Gc}^i_{t}}
|\wbf^{H}_{m}\hbf_{jk}|^{2}+\sigma^{2}}. \label{eq:update_tstar_ppp}
\end{align}
We use the same approximate closed-form beamforming solution $\wbf_{j}$
 in \eqref{eq:SGS_approx_BF} to compute \eqref{eq:update_tstar_ppp} efficiently.
This group $i^{\star}_r$ is then removed from
cluster $r$ as $\Ic_r\leftarrow\Ic_{r}\backslash \{i^{\star}_r\}$ and
is added into $\Gc_t$ as $\Gc_t\cup \{i^{\star}_r\}$.
This group assignment procedure continues until
all currently non-empty clusters have been examined for the group selection in time slot $t$.
Then, we obtain the set of scheduled groups $\Gc_t$ (and $\xbf_t$) for time slot $t$.

The above procedure repeats for $t=1,\ldots,G_{\text{max}}$
until all $\Gc_t$'s are obtained.
We summarize MGMS-GSC based on the post-processing procedure in Algorithm~\ref{alg:MGMS_hie}.

\begin{algorithm}[t]
\caption{The MGMS-GSC Algorithm for ($T, \{\xbf_t\}$)}
\label{alg:MGMS_hie}
\begin{algorithmic}[1]
\STATE \textbf{Initialization:} $t = 1$.
\STATE Obtain ($R, \{\Yc_{r}\}$) by Algorithm~\ref{alg:GSC}.
\STATE Determine $G_{\text{max}}$, $r_{\text{max}}$ from the largest cluster among all $R$ clusters.
\WHILE {$t\leq G_{\text{max}}$}
       \STATE Choose $i_{t}$ from $\Ic_{r_{\text{max}}}$ randomly.
       \STATE Update $\Ic_{r_{\text{max}}}\leftarrow\Ic_{r_{\text{max}}}\backslash \{i_{t}\}$.
       \STATE \textbf{Initialization:} Set $\Gc_t = \{i_{t}\}$, $r = 1$.
       \WHILE {$r\leq R$}
             \IF {$r\neq r_{\text{max}}$ \textbf{and} $\Yc_r\neq \emptyset$}
                 \STATE Compute $i^{\star}_r$ by \eqref{eq:update_tstar_ppp}.
                 \STATE Update $\Ic_{r}\leftarrow\Ic_{r}\backslash \{i^{\star}_r\}$,                         $\Gc_t\leftarrow\Gc_t\cup \{i^{\star}_r\}$.
                 \ENDIF
             \STATE Set $r\leftarrow r+1$.
           \ENDWHILE
       \STATE Obtain $\xbf_t$ from $\Gc_{t}$.
       \STATE Set $t \leftarrow t+1$.
    \ENDWHILE
\STATE Set $T = G_{\text{max}}$.
\RETURN \!\!($T, \{\xbf_t\}$)
\end{algorithmic}
\end{algorithm}

\section{Phase 3: Generating Multicast Beamformers}
\label{sec:phase3_MB}

Once the scheduling decision ($T, \{\xbf_t\}$)  is obtained,  in the last phase,  we solve the multi-slot
MMF multicast beamforming problem  $\Pc^{\text{bf}}_{1}(T, \{\xbf_t\})$ to determine the beamforming vector $\wbf_i$
for each group $i$.
It is straightforward to  see that this multi-slot problem   $\Pc^{\text{bf}}_{1}(T, \{\xbf_t\}$   is  separable into  $T$ per-slot multi-group MMF subproblems, where  the beamforming solution $\{\wbf_i, i\in \Gc_t\}$ is computed for the scheduled groups in $\Gc_t$ in  each time slot $t$. The per-slot MMF problem is given by
\begin{align}
\Pc^{\text{bf}}_{2,t}:\max_{\{\wbf_{i}, i\in\Gc_{t}\}} &\ \min_{i\in\Gc_{t},k\in\Kc_{i}} \, R_{ik,t}  \nn\\
&\text{s.t.} \ \sum_{i\in\Gc_{t}}\|\wbf_{i}\|^{2}\leq P.
\label{constra_transmit_power_BS}
\end{align}

Since we need to solve the above problem for each time slot,  it is desirable that the beamforming solution to $\Pc^{\text{bf}}_{2,t}$ can be computed efficiently for large-scale massive MIMO\ systems. Note that  $\Pc^{\text{bf}}_{2,t}$  is equivalent to the per-slot MMF problem $\Sc^{t}_{o}$ with the  SINR objective, shown at the beginning of Section~\ref{sec:phase1_TS_channel}. We can directly adopt the PSA-based fast algorithm, which has been discussed in Section~\ref{susbection:single_group_MMF} for the single-group MMF problem $\Sc_{1,i}$, to solves the general multi-group MMF problem $\Sc^{t}_{o}$. The algorithm is computational efficient and yields a near-optimal performance  \cite{Zhang&etal:WCL2022}.

 Finally, we point out that since both proposed scheduling algorithms, MGMS-GSS and MGMS-GSC, determine the scheduled groups for each time slot sequentially, we can solve the per-slot MMF problem $\Pc^{\text{bf}}_{2,t}$ once $\Gc_t$ is determined per time slot for transmission without delay.

\section{Simulation Results} \label{sec:simulations}

We consider a massive downlink multicast scenario
with $G=25$ groups and  $K_{i} = 5$ users/group, $i\in\Gc$
in a cell with radius $R=1~\text{km}$.
We set the receiver noise variance as $\sigma^{2} = 1$
and the BS transmit power over receiver noise as
$P/\sigma^2=10~\text{dB}$.
The user channels are generated independently
as $\hbf_{ik}\sim\mathcal{CN}({\bf{0}},\beta_{ik}{\bf{I}})$,
$k\in\Kc_{i}, i\in\Gc$,
where $\beta_{ik}$ is the user channel variance.
We model $\beta_{ik}$ by the pathloss model
$\beta_{ik} = \xi_{o}d^{-3}_{ik}$,
where the pathloss exponent is 3,
$\xi_{o}$ is the pathloss constant,
and $d_{ik}$ is the distance between the BS and user $k$ in group $i$.
We set $\xi_{o}$ such that  the nominal average received
SNR (by a single transmit antenna with unit transmit power)
at the cell boundary is $\xi_{o}R^{-3}/\sigma^2=-5~\text{dB}$.
We randomly generate user locations with $\{d_{ik}\}$ follow a uniform distribution in the range of $0.02\sim 1.0~\text{km}$.
The simulation results are averaged over
20 drops of user locations
and 20 channel realizations per user drop.

We evaluate our proposed three-phase algorithm  in Algorithm~\ref{alg:joint} for joint  group scheduling and multicast beamforming.
For comparison of different group scheduling strategies, we consider the following approaches:

\begin{itemize}
\item {\bf{MGMS-GSS}}: Algorithm~\ref{alg:joint} where Phase 2  uses MGMS-GSS by Algorithm~\ref{alg:MGMS_general}; the optimization problems in Phases 1 and 3 are solved by the PSA-based algorithm.

\item {\bf{MGMS-GSC}}: Similar to MGMS-GSS, except that MGMS-GSC by Algorithm~\ref{alg:MGMS_hie} is used in Phase 2.

\item {\bf{Single-Slot}}:   All $G$ groups are scheduled in a single time slot as the conventional multi-group multicast beamforming without scheduling, solved by the PSA-based algorithm.

\item {\bf{$\boldsymbol{G}$-Slots}}:   One group is scheduled in
each time slot with a total of $G$ time slots. The single-group multicast beamforming in each time slot is  solved by the PSA-based algorithm.
\end{itemize}

\begin{figure}[t]
\centering
\includegraphics[width=0.8\columnwidth]{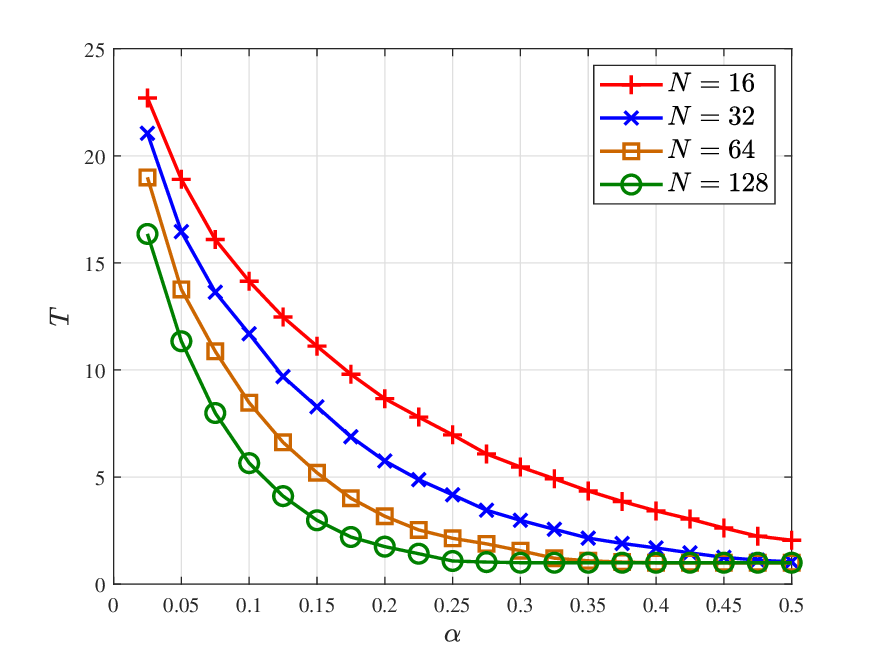}
\caption{MGMS-GSS:  Average number of  time slots $T$ vs. semi-orthogonality threshold $\alpha$.}
\label{Fig1:SGS_num_time_slots}
\centering
\includegraphics[width=0.8\columnwidth]{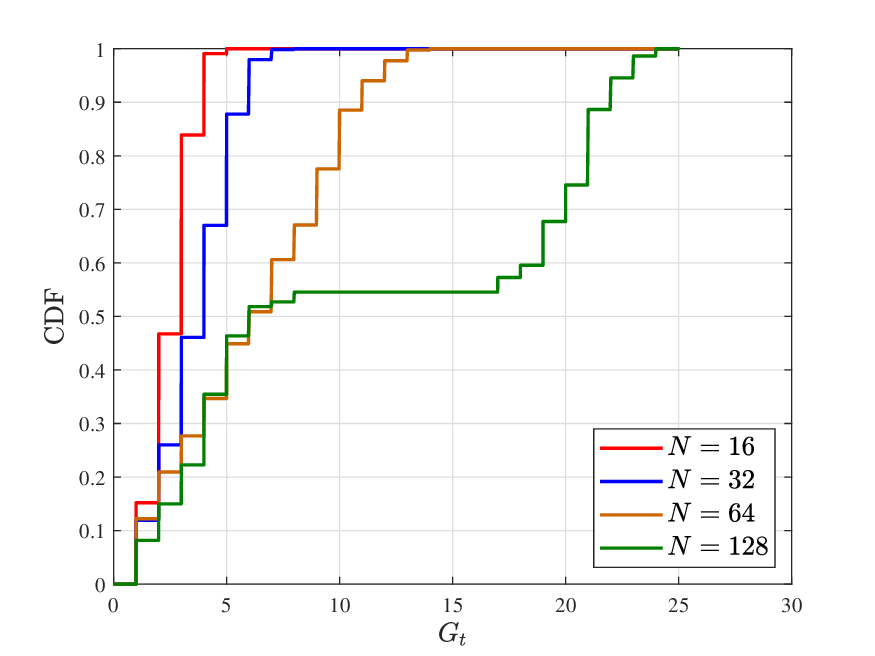}
\caption{MGMS-GSS: CDF of the number of groups $G_t$ per time slot ($\alpha=0.2$).}
\label{Fig2:SGS_cdf_ite}
\end{figure}

\subsection{Scheduling Results of MGMS-GSS }
\label{subsec:schedu_results}
We study the scheduling results of MGMS-GSS. Fig.~\ref{Fig1:SGS_num_time_slots}
shows the average number of scheduled time slots $T$
vs. the semi-orthogonality threshold $\alpha$ used in  \eqref{eq:SGS_update}, for different numbers of antennas $N$. We see that $T$ decreases as threshold $\alpha$   becomes larger.
This is expected as a larger value of $\alpha$ means a more relaxed threshold for $\hat{\hbf}_i$'s to satisfy  semi-orthogonality. Thus,  more groups will be selected into the same time slot, reducing the number of time slots required for scheduling $G$ groups. Furthermore, we observe that for the same value of $\alpha$, $T$ decreases as  $N$ becomes larger. This is because as $N$ increases, the number of degrees of freedom increases and the beam resolution becomes higher. As a result, more   groups can satisfy the semi-orthogonality criterion  and are scheduled into the same time slot, without increasing the inter-group interference.
The statistics of
the number of  scheduled groups $G_t$  per time slot are shown in Fig.~\ref{Fig2:SGS_cdf_ite}, where we plot the cumulative distribution
function (CDF) of  $G_t$  per time slot obtained by GSS (Algorithm~\ref{alg:SGS}), for different values of $N$. We set the semi-orthogonality threshold $\alpha=0.2$. We see that the CDF curves shift to the right, indicating more groups are scheduled  in a time slot as  $N$ increases, which is consistent with the observation in Fig.~\ref{Fig1:SGS_num_time_slots}. These results show that our proposed GSS in  Algorithm~\ref{alg:SGS} can  capture the level of spatial separation  among groups to effectively schedule groups  in each time slot while  maintaining a low interference level.

\begin{figure}[t]
\centering
\includegraphics[width=0.8\columnwidth]{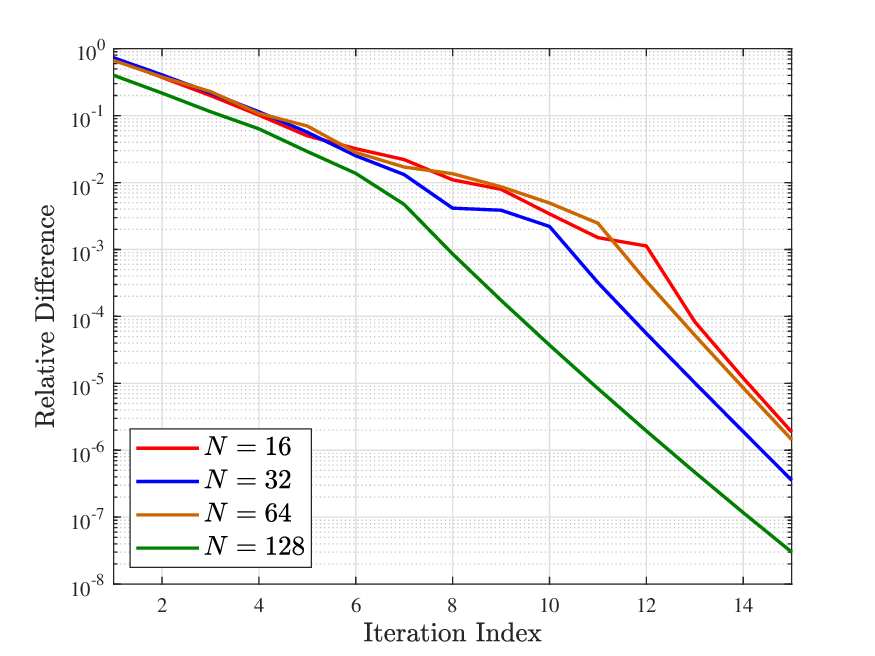}
\caption{Convergence behavior of GSC (Algorithm~\ref{alg:GSC}): Relative difference $\|\cbf^{(l+1)}_r - \cbf^{(l)}_r\|$ vs. the
iterations for cluster $1$ ($\tau=0.7$).}
\label{Fig3:GSC_inner_RD}
\centering
\includegraphics[width=0.8\columnwidth]{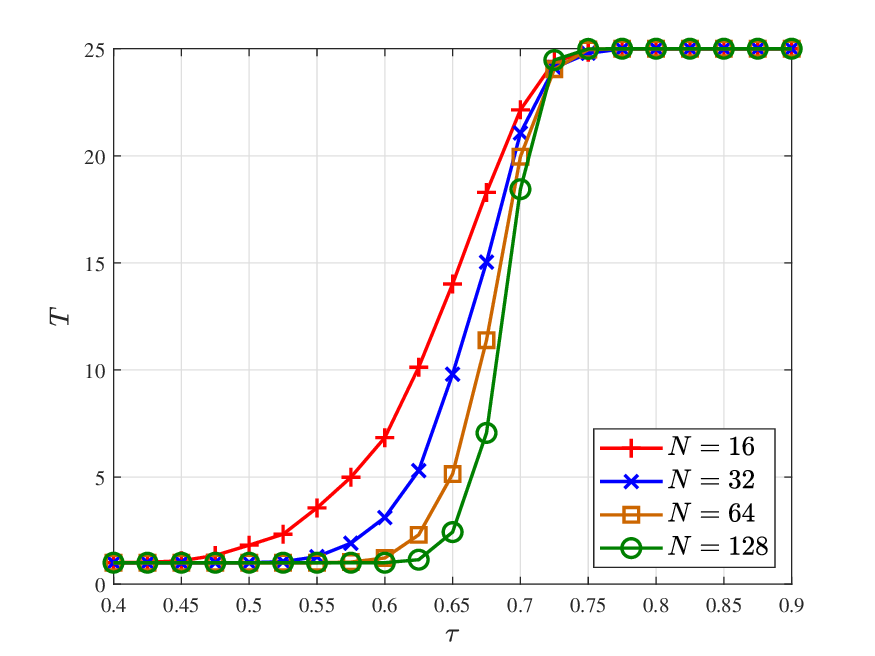}
\caption{MGMS-GSC:
 Average number of scheduled time slots $T$ vs. $\tau$.}
\label{Fig4:GSC_num_time_slots}
\end{figure}

\subsection{Scheduling Results of MGMS-GSC}
 MGMS-GSC uses  GSC (Algorithm~\ref{alg:GSC}) for clustering the groups. Note that GSC forms multiple clusters sequentially, where  each cluster $r$ is formed by updating the  centroid $\cbf_r$
iteratively until convergence. We first study the convergence behavior of GSC by Algorithm~\ref{alg:GSC}. Fig.~\ref{Fig3:GSC_inner_RD} shows the relative difference $\|\cbf^{(l+1)}_r - \cbf^{(l)}_r\|$ of the centroid in two consecutive iterations to form cluster $r=1$, for different values of $N$. We set the similarity threshold $\tau$ in \eqref{eq_GSC_Yr} to be $\tau = 0.7$.
We see that the relative difference  converges fast and drops below $10^{-3}$ within $13$ iterations. Also, the convergence speed is slightly faster as $N$ increases. This is because as $N$ increases, the degree of freedom increases. This leads to a more separable  data distribution in the dataset based on $\hat{\hbf}_i$'s, and thus, it is faster to determine the local maxima for clustering.  For the rest of simulation, we set the convergence threshold of GSC as $\|\cbf^{(l+1)}_r - \cbf^{(l)}_r\|\leq 10^{-3}$.

We now show the scheduling results of MGMS-GSC. Fig.~\ref{Fig4:GSC_num_time_slots}
plots the average number of scheduled time slots $T$ vs. similarity threshold $\tau$ used in \eqref{eq_GSC_Yr}, for different values of $N$.
We see that larger $\tau$ leads to larger $T$.
This is because larger $\tau$ leads to a bigger cluster with more groups to be considered as spatially correlated. By the final post-processing procedure, these groups in a cluster will need to be scheduled into different time slots, leading to larger $T$.
In particular, for $\tau < 0.45$, each group becomes an individual cluster, which means all groups can be scheduled into the same time slot, \ie $T=1$. \begin{figure}[t]
\centering
\includegraphics[width=0.8\columnwidth]{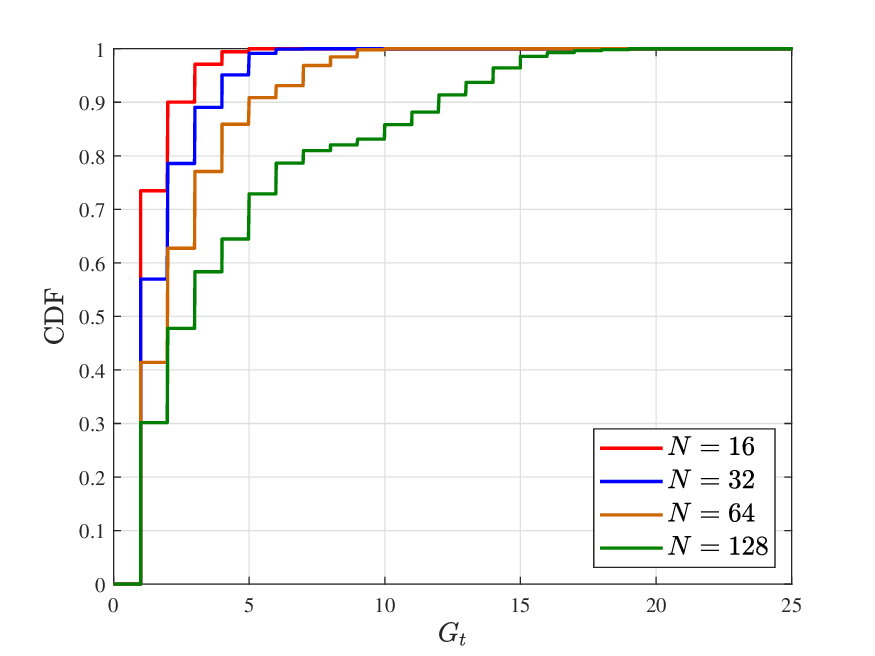}
\caption{MGMS-GSC:  CDF of
the number of groups $G_t$ per time slot ($\tau=0.7$).}
\label{Fig5:GSC_cdf_outer_ite}
\end{figure}
This becomes the conventional  Single-Slot case where all groups are scheduled
for transmission in a single time slot. For $\tau > 0.8$, a single cluster containing all groups is formed, and by the post-processing procedure, the groups are scheduled into different time slots, and we have $T=G$,
\ie one group is scheduled in each time slot. The becomes the considered $G$-Slots case.
Furthermore, for the same value of $\tau$, $T$ reduces as $N$ increases. The reason is similar to that for MGMS-GSS, \ie the degree of freedom increases
as $N$ increases, resulting in that more groups can be scheduled into the same time slot.

Fig.~\ref{Fig5:GSC_cdf_outer_ite} shows the CDF curves of
the number of scheduled groups $G_t$ per time slot, for different values of  $N$.
We set $\tau=0.7$. Similar to Fig.~\ref{Fig2:SGS_cdf_ite}, we see that as $N$ increases, $G_t$ tends to be larger, and the right tail of the CDF curve shifts to the right. This is consistent with  Fig.~\ref{Fig4:GSC_num_time_slots} with reduced $T$ as $N$ increases, as more groups are scheduled  in a time slot. Overall, we see that MGMS-GSC can capture the spatial correlation among groups to separate them  into different time slots to maintain a low interference level.

\subsection{Minimum User Throughput  Comparison}

\begin{figure}[t]
\centering
\includegraphics[width=0.8\columnwidth]{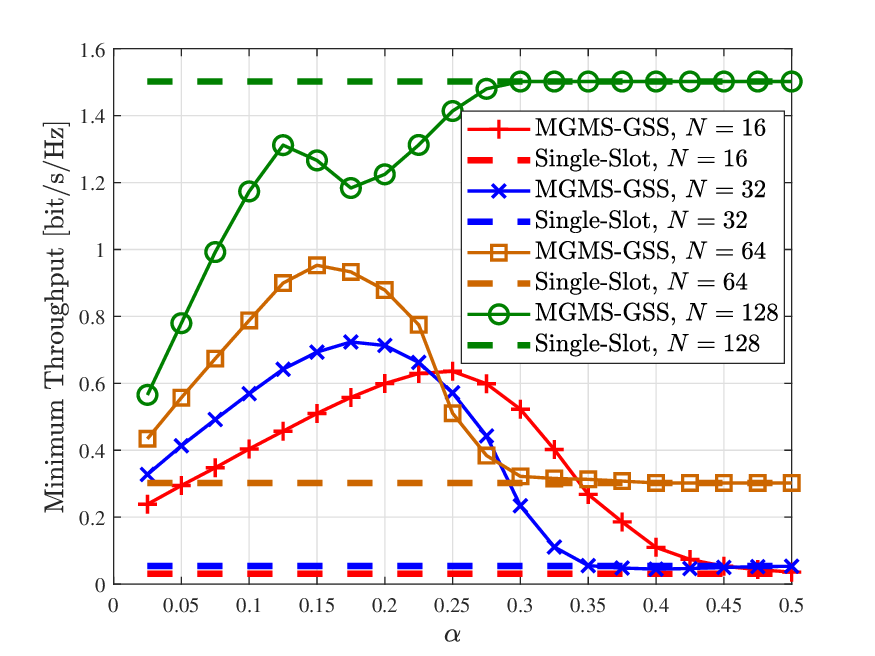}
\caption{Average minimum user throughput vs. $\alpha$.}
\label{Fig6:MGMS_GSS_min_rate}
\vspace*{1em}
\centering
\includegraphics[width=0.8\columnwidth]{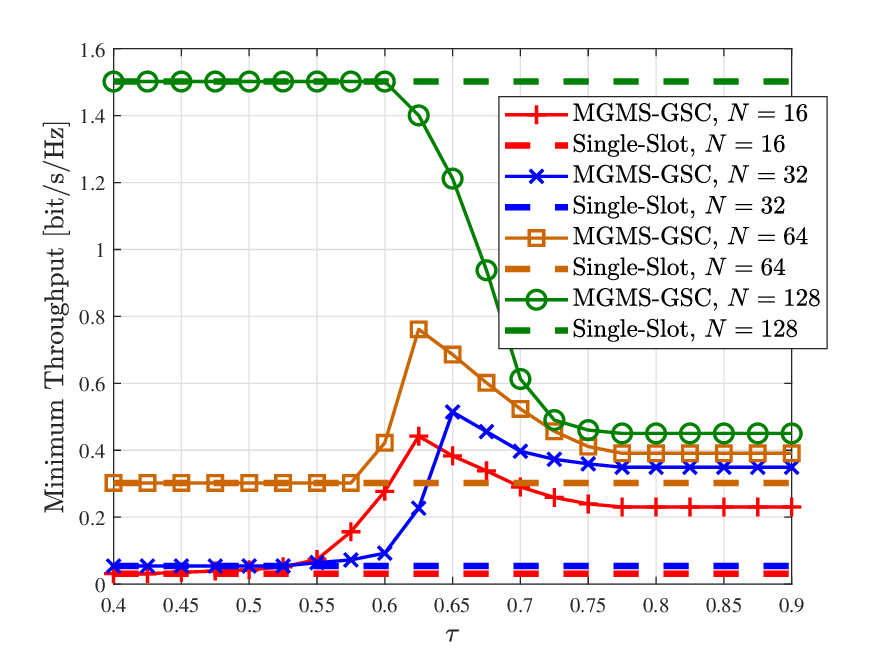}
\caption{Average minimum user throughput vs. $\tau$.}
\label{Fig7:MGMS_GSC_min_rate}
\end{figure}

We now compare the objective value of  $\Pc_{o}$, \ie the   minimum user throughput, achieved by the proposed three-phase optimization approach with different scheduling algorithms used in Phase 2.
Fig.~\ref{Fig6:MGMS_GSS_min_rate} plots
the minimum user throughput by MGMS-GSS and the benchmark method Single-Slot
over  threshold $\alpha$, for different values of $N$.
We see that for $N\leq 64$, MGMS-GSS schedules  groups in multiple time slots and achieves  higher
user throughput   than Single-Slot.
The optimal $\alpha^{\star}$ that provides the highest minimum throughput is $\alpha^\star \in [0.15,0.3]$. For $N=128$,  the optimal  $\alpha^{\star} > 0.3$,
and in this case, MGMS-GSS schedules all groups in one time slot, \ie it is identical to  Single-Slot. Intuitively, as $N$ becomes large, there are sufficient degrees of freedom to separate groups
in the spatial domain without creating much inter-group interference.
Then, scheduling all groups in one time slot can maximize the user throughput.

Fig.~\ref{Fig7:MGMS_GSC_min_rate} plots
the minimum user throughput by MGMS-GSC and Single-Slot
over threshold $\tau$.
Similar to MGMS-GSS,
MGMS-GSC schedules groups in to multiple time slots and achieves higher user throughput than
Single-Slot for $N\leq 64$
and becomes equivalent to Single-Slot for $N=128$.
The optimal $\tau^{\star}$
for the highest throughput is
$\tau^\star \in [0.6,0.7]$ for $N\leq 64$
and $\tau^\star <0.6$ for $N= 128$.
Again, for sufficiently large $N$, the minimum user throughput can be
maximized by scheduling all groups in a single time slot.

\begin{figure}[t]
\centering
\includegraphics[width=0.8\columnwidth]{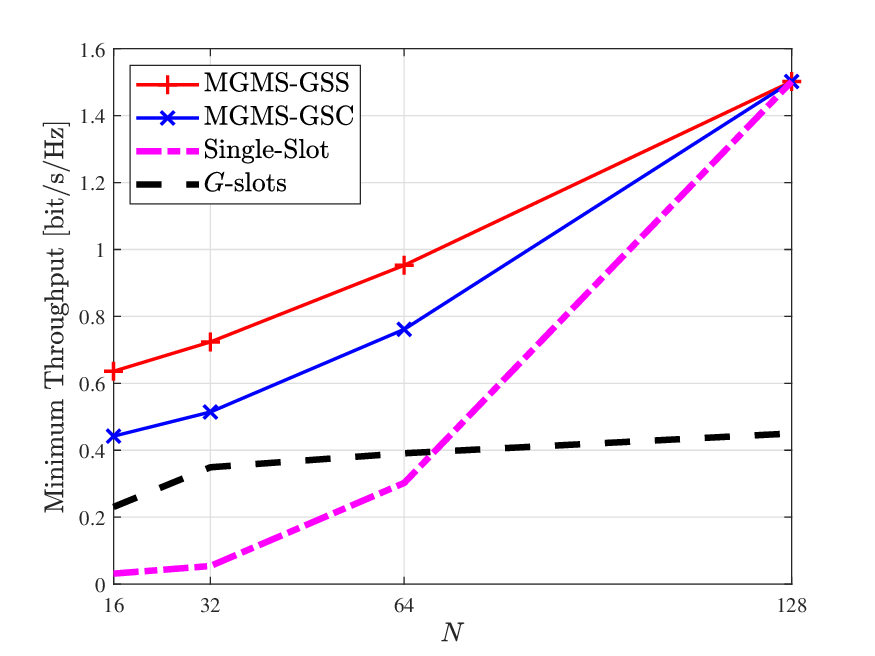}
\caption{Average minimum user throughput using optimal $\alpha^{\star}$ or $\tau^{\star}$ vs. $N$.}
\label{Fig8:Comparison_optimal_rate}
\end{figure}

We now compare the performance of different algorithms.
Fig.~\ref{Fig8:Comparison_optimal_rate} plots the average minimum
user throughput vs. the number of antennas $N$.\  The optimal threshold $\alpha^{\star}$
for MGMS-GSS and $\tau^{\star}$  for MGMS-GSC are used.
We see that both MGMS-GSS and MGMS-GSC outperform Single-Slot and $G$-Slots, demonstrating that the two algorithms can  capture the level of spatial separation among groups and make a scheduling decision effectively to improve the user throughput.  Between the two algorithms, MGMS-GSS achieves a higher throughput than  MGMS-GSC.
Note that when $N=128$, the number of antennas and users are about the same, and there are sufficient degrees-of-freedom to separate groups in the spatial domain. Thus, the optimal scheduling decision coincides with Single-Slot, \ie all groups are served simultaneously.

To compare the complexity involved in the two proposed scheduling algorithms, \ie Algorithm~\ref{alg:MGMS_general} for MGMS-GSS and Algorithm~\ref{alg:MGMS_hie} for MGMS-GSC, we show their
computation time in Table~\ref{Table1:Compt_time_N}  over different values of $N$.
We observe that both scheduling algorithms
have low computational complexity. The computation time of
MGMS-GSC only increases mildly as $N$ increases, while that of MGMS-GSS increases more noticeably.
For $N=128$, the average computation time of MGMS-GSC is $\sim 8 \%$ of that of MGMS-GSS.
Thus, MGMS-GSC is  more scalable than  MGMS-GSS.

In summary, both  MGMS-GSS and MGMS-GSC are
effective group scheduling algorithms to facilitate the multicast beamforming to maximize the minimum user throughput.
MGMS-GSS achieves higher user throughput than MGMS-GSC,
while MGMS-GSC has lower computational complexity and is more scalable than MGMS-GSS.

\begin{table}[t]
\footnotesize
\renewcommand{\arraystretch}{0}
\caption{Average Computation Time Using Optimal $\alpha^{\star}$ or $\tau^{\star}$ over $N$ (sec.)}
\label{Table1:Compt_time_N}
\centering
\begin{tabular}{l| cccccc}
\toprule
\hspace{3em}$N$                    &   16  &  32  &  64  & 128              \\ \midrule\midrule
Algorithm~\ref{alg:MGMS_general} (MGMS-GSS)    &  0.147 & 0.168 & 0.354 & 4.378    \\ \midrule
Algorithm~\ref{alg:MGMS_hie} (MGMS-GSC)    &  0.072 & 0.093 & 0.214 & 0.357    \\
 \bottomrule
\end{tabular}
\end{table}

\section{Conclusion}
\label{sec:conclusion}

This paper considers group scheduling with multicast beamforming for downlink  multicast services with many active groups. We propose a three-phase approach to the joint scheduling and beamforming optimization problem to maximize the minimum user throughput.   To determine the potential inter-group interference, we propose to use the group-channel direction of each group extracted from the optimal multicast beamforming structure for scheduling. We then propose two low-complexity group scheduling methods, MGMS-GSS and MGMS-GSC. Both two methods  utilize the group-channel direction of each group as its spatial signature but in opposite ways. MGMS-GSS measures the level of spatial separation among groups to determine a subset of groups in each time slot, while MGMS-GSC first clusters groups based on their spatial correlation and then assign  groups from different cluster to the same time slot to maximize the minimum user rate. Both MGMS-GSS and MGMS-GSC determine the  number of required time slots automatically and schedule a subset of groups in each time slot sequentially. Finally, the multicast beamformers for the scheduled groups  are efficiently computed in each time slot, by using the optimal beamforming structure with fast  PSA-based algorithm.
 Simulation results show that MGMS-GSS and MGMS-GSC can effectively explore the available spatial dimension for group scheduling to improve the minimum
user throughput.
It also shows that while MGMS-GSS achieves a higher minimum user throughput,
MGMS-GSC is a faster and more scalable approach than MGMS-GSS.

\balance
\bibliographystyle{IEEEtran}
\bibliography{Refs}

\begin{thebibliography}{10}
\providecommand{\url}[1]{#1}
\csname url@samestyle\endcsname
\providecommand{\newblock}{\relax}
\providecommand{\bibinfo}[2]{#2}
\providecommand{\BIBentrySTDinterwordspacing}{\spaceskip=0pt\relax}
\providecommand{\BIBentryALTinterwordstretchfactor}{4}
\providecommand{\BIBentryALTinterwordspacing}{\spaceskip=\fontdimen2\font plus
\BIBentryALTinterwordstretchfactor\fontdimen3\font minus
  \fontdimen4\font\relax}
\providecommand{\BIBforeignlanguage}[2]{{%
\expandafter\ifx\csname l@#1\endcsname\relax
\typeout{** WARNING: IEEEtran.bst: No hyphenation pattern has been}%
\typeout{** loaded for the language `#1'. Using the pattern for}%
\typeout{** the default language instead.}%
\else
\language=\csname l@#1\endcsname
\fi
#2}}
\providecommand{\BIBdecl}{\relax}
\BIBdecl

\bibitem{Araniti&etal:Netw2017}
G.~{Araniti}, M.~{Condoluci}, P.~{Scopelliti}, A.~{Molinaro}, and A.~{Iera},
  ``Multicasting over emerging 5{G} networks: {C}hallenges and perspectives,''
  \emph{IEEE Netw.}, vol.~31, no.~2, pp. 80--89, {M}ar./{A}pr. 2017.

\bibitem{Larsson&Edfors&Tufvesson&Marzetta:ICM:14}
E.~G. Larsson, O.~Edfors, F.~Tufvesson, and T.~L. Marzetta, ``Massive {MIMO}
  for next generation wireless systems,'' \emph{{IEEE} Commun. Mag.}, vol.~52,
  no.~2, pp. 186--195, Feb. 2014.

\bibitem{Sidiropoulos&etal:TSP2006}
N.~D. {Sidiropoulos}, T.~N. {Davidson}, and Z.-Q. {Luo}, ``Transmit beamforming
  for physical-layer multicasting,'' \emph{{IEEE} Trans. Signal Process.},
  vol.~54, no.~6, pp. 2239--2251, Jun. 2006.

\bibitem{Karipidis&etal:TSP2008}
E.~{Karipidis}, N.~D. {Sidiropoulos}, and Z.-Q. {Luo}, ``Quality of service and
  max-min fair transmit beamforming to multiple cochannel multicast groups,''
  \emph{{IEEE} Trans. Signal Process.}, vol.~56, no.~3, pp. 1268--1279, Mar.
  2008.

\bibitem{Ottersten&etal:TSP14}
D.~Christopoulos, S.~Chatzinotas, and B.~Ottersten, ``Weighted fair multicast
  multigroup beamforming under per-antenna power constraints,'' \emph{{IEEE}
  Trans. Signal Process.}, vol.~62, no.~19, pp. 5132--5142, Oct. 2014.

\bibitem{Xiang&Tao&Wang:IJWC:13}
Z.~Xiang, M.~Tao, and X.~Wang, ``Coordinated multicast beamforming in multicell
  networks,'' \emph{{IEEE} Trans. Wireless Commun.}, vol.~12, no.~1, pp.
  12--21, Jan. 2013.

\bibitem{DongLiang:CAMSAP13}
M.~Dong and B.~Liang, ``Multicast relay beamforming through dual approach,'' in
  \emph{Proc. IEEE Int. Workshop Comput. Advances Multi-Sensor Adaptive
  Process.}, Dec. 2013, pp. 492--495.

\bibitem{Tran&etal:SPL2014}
L.~{Tran}, M.~F. {Hanif}, and M.~{Juntti}, ``A conic quadratic programming
  approach to physical layer multicasting for large-scale antenna arrays,''
  \emph{{IEEE} Signal Processing Lett.}, vol.~21, no.~1, pp. 114--117, Jan.
  2014.

\bibitem{Mehanna&etal:2015}
O.~{Mehanna}, K.~{Huang}, B.~{Gopalakrishnan}, A.~{Konar}, and N.~D.
  {Sidiropoulos}, ``Feasible point pursuit and successive approximation of
  non-convex {QCQP}s,'' \emph{{IEEE} Signal Processing Lett.}, vol.~22, no.~7,
  pp. 804--808, Jul. 2015.

\bibitem{Christopoulos&etal:SPAWC15}
D.~Christopoulos, S.~Chatzinotas, and B.~Ottersten, ``Multicast multigroup
  beamforming for per-antenna power constrained large-scale arrays,'' in
  \emph{Proc. IEEE Int. Workshop Signal Process. Advances Wireless Commun.},
  Jun. 2015, pp. 271--275.

\bibitem{EbrahimiDong:Asilomar23}
M.~Ebrahimi and M.~Dong, ``Efficient design of multi-group multicast
  beamforming via reconfigurable intelligent surface,'' in \emph{Proc. Asilomar
  Conf. Signals Syst. Comput.}, Nov. 2023, pp. 1--5.

\bibitem{Mohamadi&etal:TSP22}
N.~Mohamadi, M.~Dong, and S.~ShahbazPanahi, ``{L}ow-complexity {ADMM}-based
  algorithm for robust multi-group multicast beamforming in large-scale
  systems,'' \emph{{IEEE} Trans. Signal Process.}, vol.~70, pp. 2046--2061,
  2022.

\bibitem{Sadeghi&etal:TWC17}
M.~Sadeghi, L.~Sanguinetti, R.~Couillet, and C.~Yuen, ``Reducing the
  computational complexity of multicasting in large-scale antenna systems,''
  \emph{{IEEE} Trans. Wireless Commun.}, vol.~16, no.~5, pp. 2963--2975, May
  2017.

\bibitem{Chen&Tao:ITC2017}
E.~{Chen} and M.~{Tao}, ``A{DMM}-based fast algorithm for multi-group multicast
  beamforming in large-scale wireless systems,'' \emph{{IEEE} Trans. Commun.},
  vol.~65, no.~6, pp. 2685--2698, Jun. 2017.

\bibitem{Yu&Dong:ICASSP18}
J.~Yu and M.~Dong, ``Low-complexity weighted {MRT} multicast beamforming in
  massive {MIMO} cellular networks,'' in \emph{Proc. IEEE Int. Conf. Acoust.,
  Speech, Signal Process.,}, Apr. 2018, pp. 3849--3853.

\bibitem{Ibrahim&etal:TSP2020}
M.~S. {Ibrahim}, A.~{Konar}, and N.~D. {Sidiropoulos}, ``Fast algorithms for
  joint multicast beamforming and antenna selection in massive {MIMO},''
  \emph{{IEEE} Trans. Signal Process.}, vol.~68, pp. 1897--1909, Mar. 2020.

\bibitem{Dong&Wang:TSP2020}
M.~{Dong} and Q.~{Wang}, ``Multi-group multicast beamforming: Optimal structure
  and efficient algorithms,'' \emph{{IEEE} Trans. Signal Process.}, vol.~68,
  pp. 3738--3753, May 2020.

\bibitem{Zhang&etal:WCL2022}
C.~{Zhang}, M.~{Dong}, and B.~{Liang}, ``Fast first-order algorithm for
  large-scale max-min fair multi-group multicast beamforming,'' \emph{IEEE
  Wireless Commun. Lett.}, vol.~11, no.~8, pp. 1560--1564, Aug. 2022.

\bibitem{Zhang&etal:TSP2023}
------, ``Ultra-low-complexity algorithms with structurally optimal multi-group
  multicast beamforming in large-scale systems,'' \emph{IEEE Trans. Signal
  Process.}, pp. 1--15, 2023.

\bibitem{Shadi&etal:TSP2022}
S.~{Mohammadi}, M.~{Dong}, and S.~{ShahbazPanahi}, ``Fast algorithm for joint
  unicast and multicast beamforming for large-scale massive {MIMO},''
  \emph{IEEE Trans. Signal Process.}, vol.~70, pp. 5413--5428, Oct. 2022.

\bibitem{Yoo&Goldsmith:2006JSAC}
T.~{Yoo} and A.~{Goldsmith}, ``On the optimality of multiantenna broadcast
  scheduling using zero-forcing beamforming,'' \emph{{IEEE} J. Sel. Areas
  Commun.}, vol.~24, no.~3, pp. 528--541, Mar. 2006.

\bibitem{Zhang&etal:TWC2012Scheduling}
X.~{Zhang}, M.~{Peng}, Z.~{Ding}, and W.~{Wang}, ``Multi-user scheduling for
  network coded two-way relay channel in cellular systems,'' \emph{IEEE Trans.
  Wireless Commun.}, vol.~11, no.~7, pp. 2542--2551, Jul. 2012.

\bibitem{Femenias&etal:TCOMM2016}
G.~{Femenias} and F.~{Riera-Palou}, ``Scheduling and resource allocation in
  downlink multiuser {MIMO}-{OFDMA} systems,'' \emph{IEEE Trans. Commun.},
  vol.~64, no.~5, pp. 2019--2034, May 2016.

\bibitem{Zhang&etal:TWC2017SumRate}
C.~{Zhang}, Y.~{Huang}, Y.~{Jing}, S.~{Jin}, and L.~{Yang}, ``Sum-rate analysis
  for massive {MIMO} downlink with joint statistical beamforming and user
  scheduling,'' \emph{IEEE Trans. Wireless Commun.}, vol.~16, no.~4, pp.
  2181--2194, Apr. 2017.

\bibitem{Fuchs&etal:2005}
M.~{Fuchs}, G.~{Del Galdo}, and M.~{Haardt}, ``A novel tree-based scheduling
  algorithm for the downlink of multi-user {MIMO} systems with {ZF}
  beamforming,'' in \emph{Proc. IEEE Int. Conf. Acoust., Speech, Signal
  Process.,}, vol.~3, Mar. 2005, pp. 1121--1124.

\bibitem{Razaviyayn&etal:TSP2014}
M.~{Razaviyayn}, M.~{Baligh}, A.~{Callard}, and Z.-Q. {Luo}, ``Joint user
  grouping and transceiver design in a {MIMO} interfering broadcast channel,''
  \emph{IEEE Trans. Signal Process.}, vol.~62, no.~1, pp. 85--94, Jan. 2014.

\bibitem{Nguyen&etal:Access2017}
V.-D. {Nguyen}, H.~V. {Nguyen}, C.~T. {Nguyen}, and O.-S. {Shin}, ``Spectral
  efficiency of full-duplex multi-user system: {B}eamforming design, user
  grouping, and time allocation,'' \emph{IEEE Access}, vol.~5, pp. 5785--5797,
  Mar. 2017.

\bibitem{Dimic&etal:TSP2005}
G.~{Dimi{\'c}} and N.~D. {Sidiropoulos}, ``On downlink beamforming with greedy
  user selection: {P}erformance analysis and a simple new algorithm,''
  \emph{IEEE Trans. Signal Process.}, vol.~53, no.~10, pp. 3857--3868, Oct.
  2005.

\bibitem{Shen&etal:2006}
Z.~{Shen}, R.~{Chen}, J.~G. {Andrews}, R.~W. {Heath}, and B.~L. {Evans}, ``Low
  complexity user selection algorithms for multiuser {MIMO} systems with block
  diagonalization,'' \emph{IEEE Trans. Signal Process.}, vol.~54, no.~9, pp.
  3658--3663, Sep. 2006.

\bibitem{Chen&etal:TSP2008}
R.~{Chen}, Z.~{Shen}, J.~G. {Andrews}, and R.~W. {Heath}, ``Multimode
  transmission for multiuser {MIMO} systems with block diagonalization,''
  \emph{IEEE Trans. Signal Process.}, vol.~56, no.~7, pp. 3294--3302, Jul.
  2008.

\bibitem{Zhou&Tao:2015ICC}
H.~{Zhou} and M.~{Tao}, ``Joint multicast beamforming and user grouping in
  massive {MIMO} systems,'' in \emph{Proc. IEEE Int. Conf. Commun.}, Jun. 2015,
  pp. 1770--1775.

\bibitem{Hu&etal:Netw2018}
B.~{Hu}, C.~{Hua}, C.~{Chen}, and X.~{Guan}, ``User grouping and admission
  control for multi-group multicast beamforming in {MIMO} systems,''
  \emph{Wireless Netw.}, vol.~24, no.~8, pp. 2851--2866, Apr. 2018.

\bibitem{Bandi&etal:2020TWC}
A.~{Bandi}, M.~R.~B. {Shankar}, S.~{Chatzinotas}, and B.~{Ottersten}, ``Joint
  user grouping, scheduling, and precoding for multicast energy efficiency in
  multigroup multicast systems,'' \emph{IEEE Trans. Wireless Commun.}, vol.~19,
  no.~12, pp. 8195--8210, Dec. 2020.

\bibitem{Bandi&etal:TWC2022}
------, ``Joint multislot scheduling and precoding for unicast and multicast
  scenarios in multiuser {MISO} systems,'' \emph{IEEE Trans. Wireless Commun.},
  vol.~21, no.~7, pp. 5004--5018, Jul. 2022.

\bibitem{Tran&Yue:2019Globecom}
T.~X. {Tran} and G.~{Yue}, ``{GRAB}: {J}oint adaptive grouping and beamforming
  for multi-group multicast with massive {MIMO},'' in \emph{Proc. IEEE Global
  Commun. Conf.}, Dec. 2019, pp. 1--6.

\bibitem{Yue&Qi:VTC2020}
G.~{Yue} and X.-F. {Qi}, ``Adaptive grouped physical layer multicast and
  beamforming for massive {MIMO},'' in \emph{Proc. IEEE Veh. Technol. Conf.},
  Nov. 2020, pp. 1--6.

\bibitem{Fuente&etal:2021ICC}
A.~{de la Fuente}, G.~{Interdonato}, and G.~{Araniti}, ``User subgrouping in
  multicast massive {MIMO} over spatially correlated rayleigh fading
  channels,'' in \emph{Proc. IEEE Int. Conf. Commun.}, Jun. 2021, pp. 1--6.

\bibitem{Chris&etal:2015TWC}
D.~{Christopoulos}, S.~{Chatzinotas}, and B.~{Ottersten}, ``Multicast
  multigroup precoding and user scheduling for frame-based satellite
  communications,'' \emph{IEEE Trans. Wireless Commun.}, vol.~14, no.~9, pp.
  4695--4707, Sep. 2015.

\bibitem{Cheng:1995}
Y.~{Cheng}, ``Mean shift, mode seeking, and clustering,'' \emph{IEEE Trans.
  Pattern Anal. Mach. Intell.}, vol.~17, no.~8, pp. 790--799, Aug. 1995.

\bibitem{Comaniciu&Meer:2002}
D.~{Comaniciu} and P.~{Meer}, ``Mean shift: {A} robust approach toward feature
  space analysis,'' \emph{IEEE Trans. Pattern Anal. Mach. Intell.}, vol.~24,
  no.~5, pp. 603--619, May 2002.

\end{thebibliography}
\end{document}